\documentclass[twocolumn]{aastex631}
\usepackage{natbib}
\graphicspath{{./}{figures/}}

\received{}
\revised{}
\revised{}
\accepted{}
\submitjournal{ApJ}

\shortauthors{Hasegawa et al.}

\begin{document}

\title{Spectral evolution of dark asteroid surfaces induced by space weathering over a decade}

\correspondingauthor{Sunao Hasegawa}
\email{hasehase@isas.jaxa.jp}

\author[0000-0001-6366-2608]{Sunao Hasegawa}
\affiliation{Institute of Space and Astronautical Science, Japan Aerospace Exploration Agency, 3-1-1 Yoshinodai, Chuo-ku, Sagamihara, Kanagawa 252-5210, Japan}

\author[0000-0002-8397-4219]{Francesca E. DeMeo}
\affiliation{Department of Earth, Atmospheric and Planetary Sciences, MIT, 77 Massachusetts Avenue, Cambridge, MA 02139, USA}

\author[0000-0001-8617-2425]{Micha\"{e}l Marsset}
\affiliation{Department of Earth, Atmospheric and Planetary Sciences, MIT, 77 Massachusetts Avenue, Cambridge, MA 02139, USA}
\affiliation{European Southern Observatory (ESO), Alonso de C\'{o}rdova 3107, 1900 Casilla Vitacura, Santiago, Chile}

\author[0000-0002-2934-3723]{Josef Hanu\v{s}}
\affiliation{Charles University, Faculty of Mathematics and Physics, Institute of Astronomy, V Hole\v sovi\v ck\'ach 2, CZ-18000, Prague 8, Czech Republic}

\author[0000-0001-8228-8789]{Chrysa Avdellidou}
\affiliation{Universit\'{e} C\^{o}te d'Azur, CNRS--Lagrange, Observatoire de la C\^{o}te d'Azur, CS 34229 -- F 06304 NICE Cedex 4, France}

\author[0000-0002-8963-2404]{Marco Delbo}
\affiliation{Universit\'{e} C\^{o}te d'Azur, CNRS--Lagrange, Observatoire de la C\^{o}te d'Azur, CS 34229 -- F 06304 NICE Cedex 4, France}

\author[0000-0003-4191-6536]{Schelte J. Bus}
\affiliation{Institute for Astronomy, University of Hawaii, 2860 Woodlawn Drive, Honolulu, HI 96822-1839, USA}

\author[0000-0001-8221-6048]{Hidekazu Hanayama}
\affiliation{Ishigakijima Astronomical Observatory, National Astronomical Observatory of Japan, 1024-1 Arakawa, Ishigaki, Okinawa 907-0024, Japan}

\author[0000-0001-5925-3350]{Takashi Horiuchi}
\affiliation{Institute of Astronomy, the University of Tokyo, 2-21-1 Osawa, Mitaka, Tokyo 181-0015, Japan}

\author[0000-0003-4942-2741]{Driss Takir}
\affiliation{Jacobs, NASA Johnson Space Center, Houston, TX 77058, USA}


\author[0000-0001-8923-488X]{Emmanu\"{e}l Jehin}
\affiliation{Space sciences, Technologies and Astrophysics Research Institute, Universit\'{e} de Li{e}ge, all\'{e}e du 6 Ao\^{u}t 17, 4000 Li\`{e}ge, Belgium }

\author[0000-0002-0535-652X]{Marin Ferrais}
\affiliation{Aix Marseille Universit\'{e}, CNRS, CNES, Laboratoire d'Astrophysique de Marseille, Marseille, France}

\author[0000-0002-3291-4056]{Jooyeon Geem}
\affiliation{Department of Physics and Astronomy, Seoul National University, Gwanak-gu, Seoul 08826, Republic of Korea}
\affiliation{SNU Astronomy Research Center, Seoul National University, Gwanak-gu, Seoul 08826, Republic of Korea}

\author[0000-0002-8537-6714]{Myungshin Im}
\affiliation{Department of Physics and Astronomy, Seoul National University, Gwanak-gu, Seoul 08826, Republic of Korea}
\affiliation{SNU Astronomy Research Center, Seoul National University, Gwanak-gu, Seoul 08826, Republic of Korea}

\author{Jinguk Seo}
\affiliation{Department of Physics and Astronomy, Seoul National University, Gwanak-gu, Seoul 08826, Republic of Korea}
\affiliation{SNU Astronomy Research Center, Seoul National University, Gwanak-gu, Seoul 08826, Republic of Korea}

\author[0000-0002-2618-1124]{Yoonsoo P. Bach}
\affiliation{Department of Physics and Astronomy, Seoul National University, Gwanak-gu, Seoul 08826, Republic of Korea}
\affiliation{SNU Astronomy Research Center, Seoul National University, Gwanak-gu, Seoul 08826, Republic of Korea}

\author[0000-0002-0460-7550]{Sunho Jin}
\affiliation{Department of Physics and Astronomy, Seoul National University, Gwanak-gu, Seoul 08826, Republic of Korea}
\affiliation{SNU Astronomy Research Center, Seoul National University, Gwanak-gu, Seoul 08826, Republic of Korea}

\author[0000-0002-7332-2479]{Masateru Ishiguro}
\affiliation{Department of Physics and Astronomy, Seoul National University, Gwanak-gu, Seoul 08826, Republic of Korea}
\affiliation{SNU Astronomy Research Center, Seoul National University, Gwanak-gu, Seoul 08826, Republic of Korea}

\author[0000-0002-7363-187X]{Daisuke Kuroda}
\affiliation{Japan Spaceguard Association, Bisei Spaceguard Center 1716-3 Okura, Bisei, Ibara, Okayama 714-1411, Japan}

\author[0000-0002-9995-7341]{Richard P. Binzel}
\affiliation{Department of Earth, Atmospheric and Planetary Sciences, MIT, 77 Massachusetts Avenue, Cambridge, MA 02139, USA}

\author[0000-0001-6990-8496]{Akiko M. Nakamura}
\affiliation{Department of Planetology, Graduate School of Science, Kobe University, 1-1 Rokkodai-cho, Nada-ku, Kobe 657-8501, Japan}

\author[0000-0002-5033-9593]{Bin Yang}
\affiliation{N\'{u}cleo de Astronom\'{i}a, Facultad de Ingenier\'{i}ay Ciencias, Universidad Diego Portales, Chile}

\author[0000-0002-2564-6743]{Pierre Vernazza}
\affiliation{Aix Marseille Universit\'{e}, CNRS, CNES, Laboratoire d'Astrophysique de Marseille, Marseille, France}




\begin{abstract} 
The surface of airless bodies like asteroids in the Solar System are known to be affected by space weathering.
Experiments simulating space weathering are essential for studying the effects of this process on meteorite samples, but the problem is that the time spent to reproduce space weathering in these experiments is billions of times shorter than the actual phenomenon.
In December 2010, the T-type asteroid 596 Scheila underwent a collision with a few-tens-of-meters impactor.
A decade later, there is an opportunity to study how the surface layer of this asteroid is being altered by space weathering after the impact.
To do so, we performed visible spectrophotometric and near-infrared spectroscopic observations of 596 Scheila.
The acquired spectrum is consistent with those observed shortly after the 2010 impact event within the observational uncertainty range.
This indicates that the surface color of dark asteroids is not noticeably changed by space weathering over a 10-year period.
This study is the first to investigate color changes due to space weathering on an actual asteroid surface in the Solar System.
Considering that fresh layers are regularly created on asteroid surfaces by collisions, we suggest a genetic link between D/T-type and dark (low albedo) X-complex asteroids and very red objects such as 269 Justitia, 732 Tjilaki (and 203 Pompeja).
New observations show that 203 Pompeja has a X-type-like surface, with some local surface areas exhibiting a very red spectrum.
\end{abstract}



\section{Introduction} \label{sec:intro}
It is known that the surface layers of airless bodies in the Solar System are usually not fresh, having been exposed to space for long periods of time.
The effect responsible for altering the surfaces of these objects and affecting their colors and reflectance spectra is called space weathering.
This process was first evidenced by the distinct colors of the youngest craters on the Moon \citep{Gold1955}.
Analysis of lunar soil samples then revealed that space weathering on the lunar surface is caused by the impact of micrometeoroids and implantation of solar wind \citep{Hapke1975}.

The lack of main-belt asteroids with spectra similar to those of ordinary chondrites suggested that space weathering also affected the surface layers of S-complex asteroids \citep[e.g.,][]{Pieters1994}.
Telescopic observations of S-complex and Q-type asteroids \citep{Binzel1996} and rendezvous observations of S-complex asteroids by spacecraft \citep[e.g., ][]{Veverka2000, Hiroi2006} later confirmed that space weathering causes changes in the reflectance spectra of the surfaces of S-complex asteroids.
In particular, the samples returned from the S-complex asteroid 25143 Itokawa were compositionally similar to ordinary chondrites \citep[e.g., ][]{Nakamura2011, Yurimoto2011}.  
Nanophase iron particles, which had been found on the lunar soil \citep{Keller1993}, were also found in the Itokawa samples \citep{Noguchi2011}, thereby proving that the reflectance spectra of S-complex asteroids are reddened by space weathering.

For a long time, space weathering of dark asteroids (albedo values less than 0.1) such as C-complex and D-type asteroids did not receive much attention because many dark asteroids had been found to exhibit spectra consistent with carbonaceous chondrites  \citep[e.g., ][]{Clark2002}.
\citet{Nesvorny2005} was the first to show a trend in space weathering of C-complex asteroids based on correlations between family ages and spectral slopes.
\citet{Lantz2018} further showed trends in spectral slope due to space weathering for each spectral type of dark asteroids observed from telescopic and laboratory study data.

Laboratory simulation experiments of space weathering using minerals and meteorites are very important for understanding spectral changes in response to this process on the surface of asteroids \citep[e.g., ][]{Moroz1996, Yamada1999, Dukes1999}.
In particular, the results of simulated space weathering experiments on carbonaceous chondrites, which are considered to be the meteorite counterparts of dark asteroids, are essential to understand the spectral changes induced by space weathering on this type of object \citep[e.g., ][]{Vernazza2013, Matsuoka2015, Lantz2017}.
While laboratory experiments are important research methods for understanding space weathering, there is one obstacle that cannot be overcome: time simulated experiment alter the spectra of the exposed samples over milli-year timescales, whereas the spectral changes induced by space weathering on actual asteroids act over thousand-year timescales or more.
Although the irradiating fluxes are increased billions of times in intensity with respect to actual space weathering during the experiments, it cannot be guaranteed that this properly reproduces what is happening in nature over such long timescales.
If experiments could be performed on meteorites such that they match the space-weathered age of the actual observed asteroid and if the resulting irradiated meteorite and the asteroid spectra could be compared, this problem would be resolved.

Another point of interest is the characteristic time over which space weathering significantly affects the spectra.
Studies combining asteroid spectra with the age of their respective collisional families showed spectral changes in response to space weathering over million- to billion-year timescales \citep{Nesvorny2005, Vernazza2009}.
In addition, space weathering has been reported to be efficient over thousand year timescales, although this was reported for ordinary chondrites instead of carbonaceous chondrites \citep{Noguchi2014,JinSubmited}.
However, it remains unknown how spectra change due to space weathering over shorter time periods.
In particular, it would be very interesting to know if spectral slope changes linearly or logarithmically with time over periods shorter than a thousand to a million years.

The surface of the T-type asteroid 596 Scheila with an albedo of 0.038\footnote{Mean values obtained from IRAS \citep{Tedesco2002},  AKARI \citep{Usui2011}, and WISE \citep{Masiero2011}.\label{pv}} was refreshed during an impact with a few-tens-of-meters projectile in 2010 \citep{Ishiguro2011a, Ishiguro2011b}.
In this collision event, the spectral type of 596 turned redder, from T type to D type \citep{Hasegawa2022}.
Assuming that the surface layer of 596 was space-weathered before the impact, and that a fresh surface appeared after the event, it was concluded that objects compositionally similar to 596 become bluer in spectral slope due to space weathering.
This year marks approximately 10 years since the collisional event.
This means that the refreshed surface of 596 has been exposed to space weathering for a period of approximately 10 years.
To test whether the surface layer of 596 and its reflectance spectrum changed during that period of time, we performed spectral observations of 596 in 2022, more than a decade after the collisional event.

\begin{figure*}
\gridline{\fig{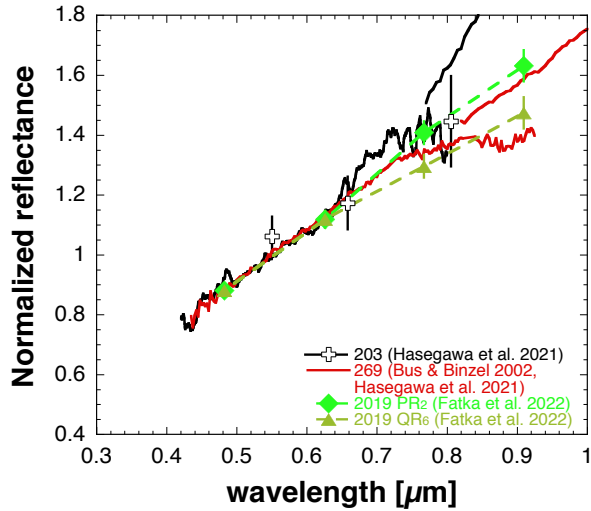}{0.5\textwidth}{}
          }
\caption{
Comparison of the visible spectra of VROs: 203 Pompeja, 269 Justitia, 2019 ${\mathrm{PR_{2}}}$, and 2019 ${\mathrm{QR_{6}}}$.
The figure shows the spectra in the 0.4--1.0 $\mu$m wavelength regions, normalized to 1 at 0.55 $\mu$m.
}
\label{fig:VRO}
\end{figure*}

In this study, we investigate the spectral changes caused by space weathering over a short timescale of 10 years, which is a much shorter timescale than the formation age of the solar system. 
A comparable study of asteroids over a short timescale of 300 years was recently reported by \citet{Fatka2022}.
These authors found that two near-Earth asteroids (NEAs) 2019 ${\mathrm{PR_{2}}}$ and 2019 ${\mathrm{QR_{6}}}$ may have formed as a pair about three hundred years ago.
The authors also showed that the two asteroids have spectral slopes that are equal to or redder than those of D-type asteroids.
This suggests that the fresh surface layers of D-type asteroids before space weathering effects become apparent may be similar to very red objects (VROs) found by \citet{Hasegawa2021b}, which have a redder spectral slope than the average D-type asteroid \citep{DeMeo2009, Hasegawa2022}.
Figure \ref{fig:VRO} shows that the very red NEAs, which may be very young, have a similar slope in visible spectra to the VROs 203 Pompeja with an albedo of 0.045\textsuperscript{\ref{pv}} and 269 Justitia with an albedo of 0.080\textsuperscript{\ref{pv}}.
This indicates that the surface ages of 203 and 269 may be as young as a few hundred years.
Therefore, we also conducted observations of 203 and 269 to investigate if a slope change occurred in their spectra over time.
Note that one further VRO: 732 Tjilaki with an albedo of 0.068\footnote{Mean values obtained from IRAS \citep{Tedesco2002} and  AKARI \citep{Usui2011}.\label{pv2}} was discovered in \citet{BourdelledeMicasinPress}.
We also investigated if a slope change of occurred in 732 spectra over time using literature. 

\section{Observations and Data Analysis} \label{sec:Observations and Data Analysis}
New near-infrared (NIR) observations of 596 Scheila were carried out with the SpeX NIR spectrograph \citep{Rayner2003} on the 3.2 m NASA Infrared Telescope Facility (IRTF) located on Maunakea, Hawaii on 2022 April 4.
Both Landolt 102-1081 and Landolt 107-998 were used as solar analogues for the NIR observation \citep{Marsset2020}.
Reduction of NIR spectral images was conducted using the Image Reduction and Analysis Facility (IRAF; \citealt{Tody1993}) and the Interactive Data Language (IDL).
Close in time, visible observations were conducted in the synchronous imaging camera with $g'$-, $R_{\rm C}$-, and $I_{\rm C}$-bands \citep{Yanagisawa2010AIPC} on the 1.05 m Murikabushi Telescope located on Ishigakijima, Japan on 2022 Febrary 8.
HD142801 was used as a calibration star for the visible observation.
Data reduction for photometry was performed using the IRAF. 
The procedures for the NIR and visible observations and reduction are as described in \citet{Binzel2019} and \citet{Hasegawa2021b}, respectively.

NIR spectrum of 269 Justitia was collected in 2009 June 15 using SpeX/IRTF (Emery private comm.).
HD 124901 was used as a solar analogue star.
Spectral images were reduced by the IDL.
The procedures for the NIR observations and reduction of 269 are as described in \citet{Emery2011}.

NIR spectra of 203 Pompeja were obtained in 2020 December 1 and 2022 June 10 and 27.
As in previous observations, the observations were made using SpeX/IRTF.
BD+00 2717, HD 249566, and HD 106116 were used as a solar analogue star on 2020 December 1, 2022 June 10, and  2022 June 27, respectively.
Reduction of the spectroscopic images was performed with the IDL.
The procedures for the NIR observations and reduction of 203 are as described in \citet{Takir2012} and \citet{Avdellidou2021}.
Visible spectra of 203 Pompeja were obtained on 2022 June 18, 19, 29, and 30.
Colorimetric observations were acquired with the robotic telescope TRAPPIST-South (TS) \citep{Jehin2011} located, at the ESO La Silla Observatory in Chile. 
TS is a 0.6 m Ritchey-Chr\'{e}tien telescope operating at f/8 and equipped with a thermoelectrically-cooled FLI ProLine 3041-BB CCD camera. 
The data were reduced using the PHOTOMETRYPIPELINE \citep{Mommert2017} and calibrated to the band they were observed using the Pan-STARRS catalogue.

\section{596 Scheila} \label{sec:596}
\subsection{Spectroscopic results} 
We compared the spectra of 596 Scheila obtained before, immediately after the 2010 impact, and a decade after the event to search for any spectral change induced by space weathering. 
Spectral data before the 2010 impact event are from \citet{Bus2002} and \citet{DeMeo2009}.
NIR spectral data obtained immediately after the impact event were retrieved from \citet{Yang2011} and \citet{Hasegawa2022}.
Visible spectrophotometric data immediately after impact were combined from \citet{Trigo-Rodriguez2011EPSC}, \citet{Betzler2012}, \citet{Hsieh2012}, \citet{Jewitt2012}, and \citet{Shevchenko2016}.
In addition to the data acquired in the present work, multi-filter photometric data from \citet{Sergeyev2022} were used as part of the ``10 years after the event'' dataset as those data were obtained on 2018 August 20.
As reference data, visible reflectance spectrum from Gaia is an average of several observations taken somewhere between 2014 August 5 and 2017 May 28, i.e., $\sim$3.5 and $\sim$6.5 years after the impact event are also shown \citep{GaiaCollaborationInPress}\footnote{The epoch spectra are not available in the Data Release 3 (DR3). 
All the DR3 asteroid spectra are averages of epoch spectra taken between 2014 August 5 and 2017 May 28.\label{JPL}}.
Table \ref{tab:1} indicates the sub-observer coordinates of NIR observations before, immediately after, and a decade after the impact, calculated based on the shape model of 596 from \citet{Hanus2021}.
Although the observed latitude and longitude of each observation is different, \citet{Hasegawa2022} indicated that the ejecta discharged by the 2010 impact event was deposited over the entire surface of the asteroid and that the surface layer was homogeneously covered by fresh material.

\begin{figure*}
\gridline{\fig{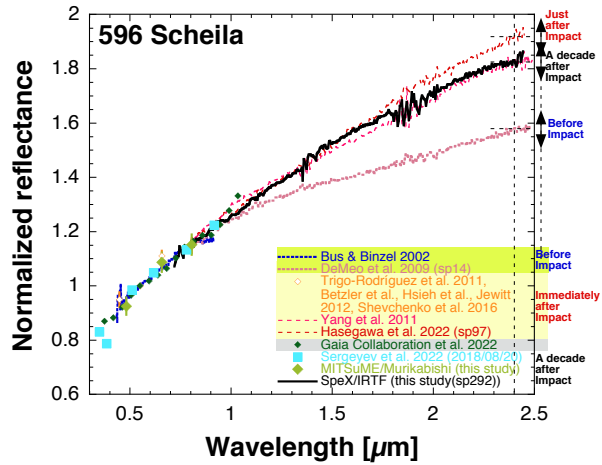}{0.5\textwidth}{}
          }
\caption{
Comparison of 596 Scheila's NIR spectra before, immediately after, and  a decade after the 2010 impact event.
The figure shows spectra of 596 in the 0.4--2.5 $\mu$m wavelength region, normalized to 1 at 0.55 $\mu$m.
The black arrows indicate the slope uncertainty from \citet{Marsset2020} centered at the reflectance value of the spectra at 2.40 $\mu$m, as shown by the black dashed lines.
}
\label{fig:spectra}
\end{figure*}

Figure \ref{fig:spectra} compares the reflectance spectrum of 596 before, immediately after, and 10 years after the impact event.
The spectral slope of 596 over 0.8--2.5 $\mu$m before, immediately after, and a decade after the impact event is 25.5, 46.7(40.5), and 42.2\% $\mu$m$^{-1}$, respectively.
The slopes of the NIR spectra immediately after the impact and about 10 years after the impact are consistent within the expected slope uncertainty of 4.2\% $\mu$m$^{-1}$ for SpeX measurements \citep{Marsset2020}.
The slope consistency between the visible spectra of 596 obtained over all observed periods indicates no significant change over the decade at these wavelengths within the observational uncertainty range.  

\begin{deluxetable*}{llccccccccl}
\label{tab:1}
\tablecaption{Sub-observer coordinates of 596 Scheila during NIR observations based on \citet{Hanus2021}}
\tablewidth{0pt}
\tablehead{
\colhead{Observation time}&\colhead{Timing of }&\multicolumn4c{Sub-Earth point}&\colhead{Spectral slope}&$R_{\rm h}^{c}$&$\Delta^{c}$&$\alpha^{c}$&\colhead{Reference}\\
\colhead{}&\colhead{Observation}&\colhead{${\lambda_{1}}^{a}$ [\degr]}&\colhead{${\phi_{1}}^{a}$ [\degr]}&\colhead{${\lambda_{2}}^{b}$ [\degr]}&\colhead{${\phi_{2}}^{b}$ [\degr]}&\colhead{[\% $\mu$m$^{-1}$]}&[au]&[au]&[\degr]&\colhead{}
}
\startdata
2002/06/01 09:40&before impact&137&46&232&$-$47&25.5&2.45&1.43&0.4&\cite{DeMeo2009}\\ 
2011/01/04 14:50&immediately after impact&132 &$-$39&263&43&$40.5^{d}$&3.07&2.26&12.0&\cite{Yang2011}\\
2011/01/05 14:20&immediately after impact&319 &$-$39&90&44&$40.5^{d}$&3.07&2.25&11.7&\cite{Yang2011}\\
2011/02/07 09:31&immediately after impact&99&$-$40&225&50&46.7&3.02&2.07&6.4&\cite{Hasegawa2022}\\ 
2022/04/09 14:29&a decade after impact&357&51&179&$-$42&42.2&2.46&1.73&19.2&This study\\ 
\enddata
\tablecomments{
\\ 
$^{a}$Pole solution 1: ${\lambda}_1$ = 110 [\degr], ${\beta}_1$ = $-$24 [\degr], Period = 15.8605 [hr]\\ 
$^{b}$Pole solution 2: ${\lambda}_2$ = 273  [\degr], ${\beta}_2$ = $-$38  [\degr], Period = 15.8609 [hr]\\
$^{c}$The heliocentric ($R_{\rm h}$) and geocentric distances ($\Delta$) and phase angle ($\alpha$) for asteroid observations were taken from the NASA/ Jet Propulsion Laboratory (JPL) HORIZONS ephemeris generator system$^{e}$.
$^{d}$The value of the combined spectrum in \citet{Yang2011}.
$^{e}$ https://ssd.jpl.nasa.gov/horizons\\
}
\end{deluxetable*}

\subsection{Spectral changes due to phase angle} 
Regarding possible changes in the spectral gradient, it was shown in \citet{Hasegawa2022} that particle size effect can not explain the observed spectral reddening changes.
However, the possibility of a change in the slope of the spectrum due to a change in roughness and phase angle was not discussed in \citet{Hasegawa2022} and is discussed below.

\citet{Binzel2015} pointed out that spectral changes occur due to surface roughness effects in case of regolith-free surfaces.
However, the surface layer of 596 is considered to be covered with regolith \citep{Ishiguro2011a,Yang2011}.
Therefore, changes in spectral slope due to roughness do not need to be taken into account in the case of 596.

\begin{figure*}
\gridline{\fig{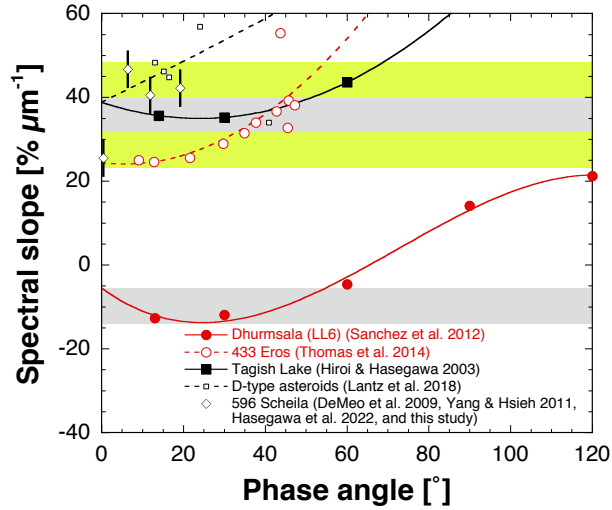}{0.5\textwidth}{}
          }
\caption{Comparison between the spectral slope of Dhurmsala (LL6) and Tagish Lake (C2) meteorites and Sw-type asteroid 433 Eros and D-type asteroids as function of phase angle.
The D-type asteroid data set is a combination of multiple D-type asteroids.
Gray and yellow areas indicate the slope uncertainty from \citet{Marsset2020} for the meteorites and the asteroids, respectively.
The solid and dashed lines represent polynomial fits for the meteorites and asteroids, respectively.
The D-type dataset includes the post-impact spectrum of 596.
}
\label{fig:slope}
\end{figure*}

\citet{Sanchez2012} described the spectral dependence on the phase angle of ordinary chondrite.
For ordinary chondrites, changes in spectral slope are not seen below a phase angle of 30 degrees (Figure \ref{fig:slope}).
Considering the observational slope uncertainty \citep{Marsset2020}, unless the observations are repeated several times, in which case the uncertainty on the mean can be reduced (Figure \ref{fig:slope}).
\citet{Thomas2014} investigated the spectral dependence on the phase angle of S-type asteroids; 433 Eros, 1036 Ganymed, and 1627 Ivar.
Among these objects, 433 Eros was shown to be spectrally homogeneous across its surface by NASA's spacecraft NEAR Shoemaker \citep{Bell2002}. 
No change in spectral slope of 433 is observed up to within 22 degrees of the phase angle  (Figure \ref{fig:slope}).
Given the uncertainty of the slope of the observations shown in \citet{Marsset2020}, unless there are multiple observations, in which case the uncertainty on the mean can be reduced (Figure \ref{fig:slope}).

While there are examples of phase angle dependence for S-type asteroids as described above, no studies have investigated the phase-angle dependency of spectral slope of individual D-type asteroids. 
However, there are examples where spectra of various D-type asteroids have been collected to investigate the dependence of the spectral slope on the phase angle in the visible to near-infrared wavelength range \citep{Lantz2018}.
D-type asteroids have been shown to have an increasing spectral slope with increasing phase angle, but, given the uncertainty in the slope of the observations presented in \citet{Marsset2020}, it is not possible to state the change in the slope of the spectrum within a phase angle between $\sim$2 and $\sim$19 degrees considering the number of available measurements (Figure \ref{fig:slope}).
However, this dataset used data from different D-type asteroids, which may have different spectral slopes at a given phase angle.
\citet{Hiroi2003} studied the spectral slope of the Tagish lake meteorite by varying the phase angle.
The slope of the spectrum does not change up to a phase angle of 30 degrees, but the slope of the spectrum increases slightly at a phase angle of 60 degrees  (Figure \ref{fig:slope}).
However, up to within $\sim$50 degrees of the phase angle, the change is within the range of observation uncertainty.
Therefore, 596 having a spectrum comparable to that of the Tagish Lake meteorite one cannot expect to detect a slope change at phase angles below $\sim$50 degrees.

NIR observations of 596 before and after the 2010 impact cover a range of phase angles comprised between 0.4 to 19.2 degrees. 
Considering that the spectral slope of 596 varies beyond about three times the observation uncertainty, it can be concluded that the spectral changes are not due to changes caused by the phase angle.

\subsection{Constraints on spectral changes due to space weathering} 
The spectral match between the spectrum of the fresh surface observed immediately after the 2010 impact event and the space-weathered one observed a decade later indicates that the spectra of dark asteroids like 596 do not evolve over a decade timescale due to space weathering.
This was predicted from a number of space weathering experiments performed on carbonaceous chondrites \citep[e.g., ][]{Vernazza2013, Matsuoka2015, Lantz2017}.
However, this is the first time that the absence of spectral changes due to space weathering is observationally confirmed for an actual asteroid under conditions where the exposure time  is precisely known.

\begin{figure*}
\gridline{\fig{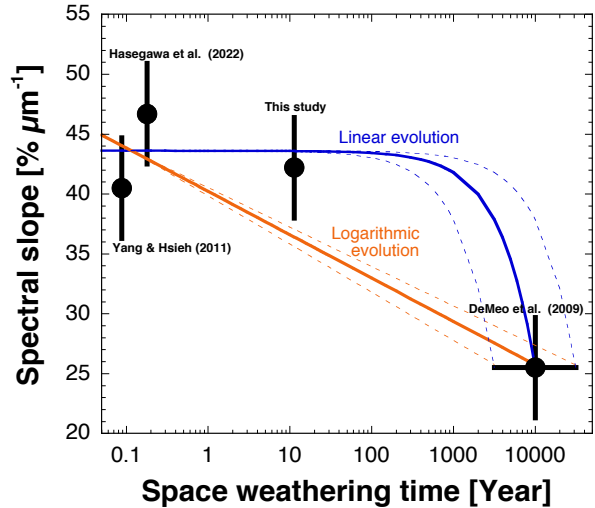}{0.5\textwidth}{}
          }
\caption{
Spectral slope for 596 Scheila against space weathering time.
Blue and orange lines correspond to linear and logarithmic regression of the spectral slope values of \citet{DeMeo2009}, \citet{Yang2011}, and \citet{Hasegawa2022}, respectively.
}
\label{fig:LinearLog}
\end{figure*}

Next, we investigate whether the spectral change due to space weathering acts linearly or logarithmically with time. 
Following \citet{Ishiguro2011b}, we assume that the impact on 596 occurred on 2010 December 3. 
Then, considering that the average collisional lifetime of 596 for impacts of the scale of the 2010 event is about 10$^{4}$ yr \citep{Hasegawa2022}, we assume that this was the age of 596's surface just before the impact, when the spectrum of \citet{DeMeo2009} was acquired. 
Under these assumptions, we fitted both a linear and a logarithmic function to the slope values of the spectra from \citet{Yang2011} (t=0.09 yr after the collision), \citet{Hasegawa2022} (t=0.18 yr), and \citet{DeMeo2009} (t=10$^{4}$ yr). 
By doing so, we find that the slope would have changed by 0.0018\% $\mu$m$^{-1}$ yr$^{-1}$ assuming a linear evolution and 3.6\% $\mu$m$^{-1}$ (log(yr))$^{-1}$ assuming a logarithmic evolution (Figure \ref{fig:LinearLog}). 
That is, if space weathering changes linearly, the expected slope after about 10 years would be 43.6\% $\mu$m$^{-1}$, while if it changes logarithmically, the expected slope after about 10 years would be 36.4\% $\mu$m$^{-1}$.
The actual slope we observed after about 10 years is 42.2\%, meaning that the change is smaller than the expected value from the logarithmic function, although 1.5-$\sigma$ consistent considering the expected 4.2 \% $\mu$m$^{-1}$ measurement uncertainty \citep{Marsset2020}.
This suggests that the spectral change slope evolution due to space weathering during the first decade is not logarithmic with respect to time, but linear.
Therefore, we find that the NIR reflectance spectra of reddish dark asteroids like 596 likely do not evolve faster than a logarithmic function, at least during the first 10 years following a refreshing event.

\section{269 Justitia} \label{sec:269}
\begin{figure*}
\gridline{\fig{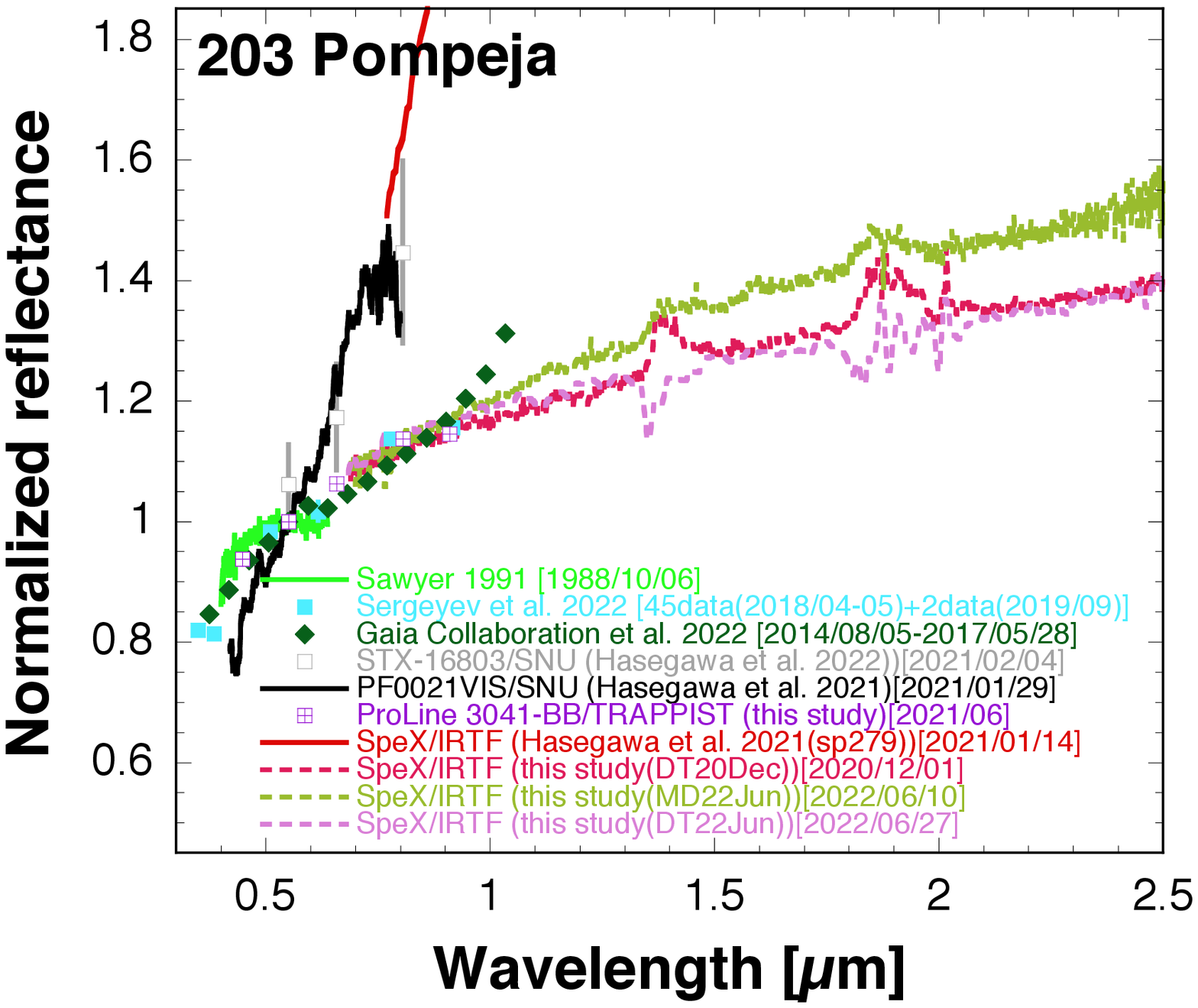}{0.5\textwidth}{}
             \fig{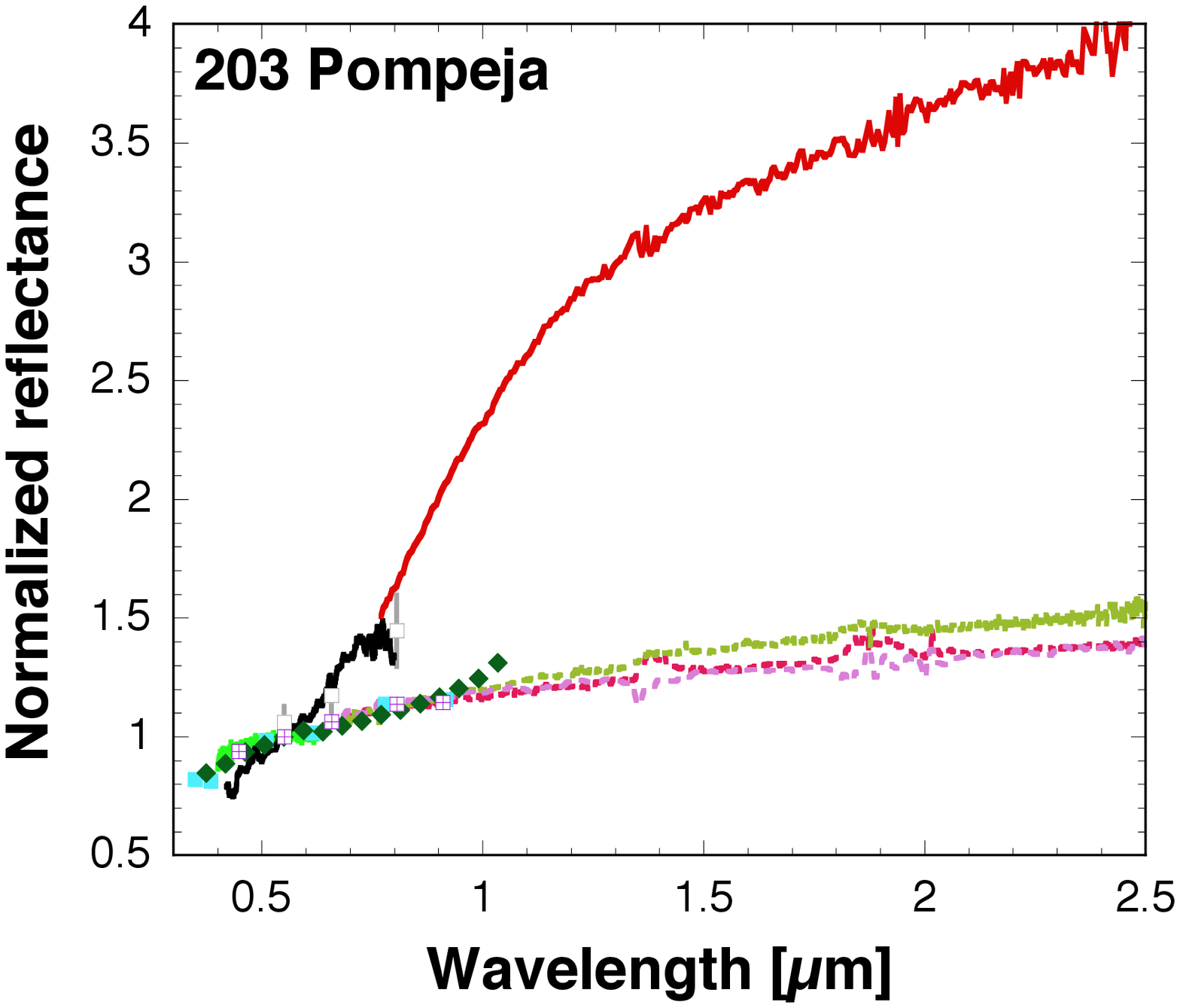}{0.5\textwidth}{}
}
\gridline{\fig{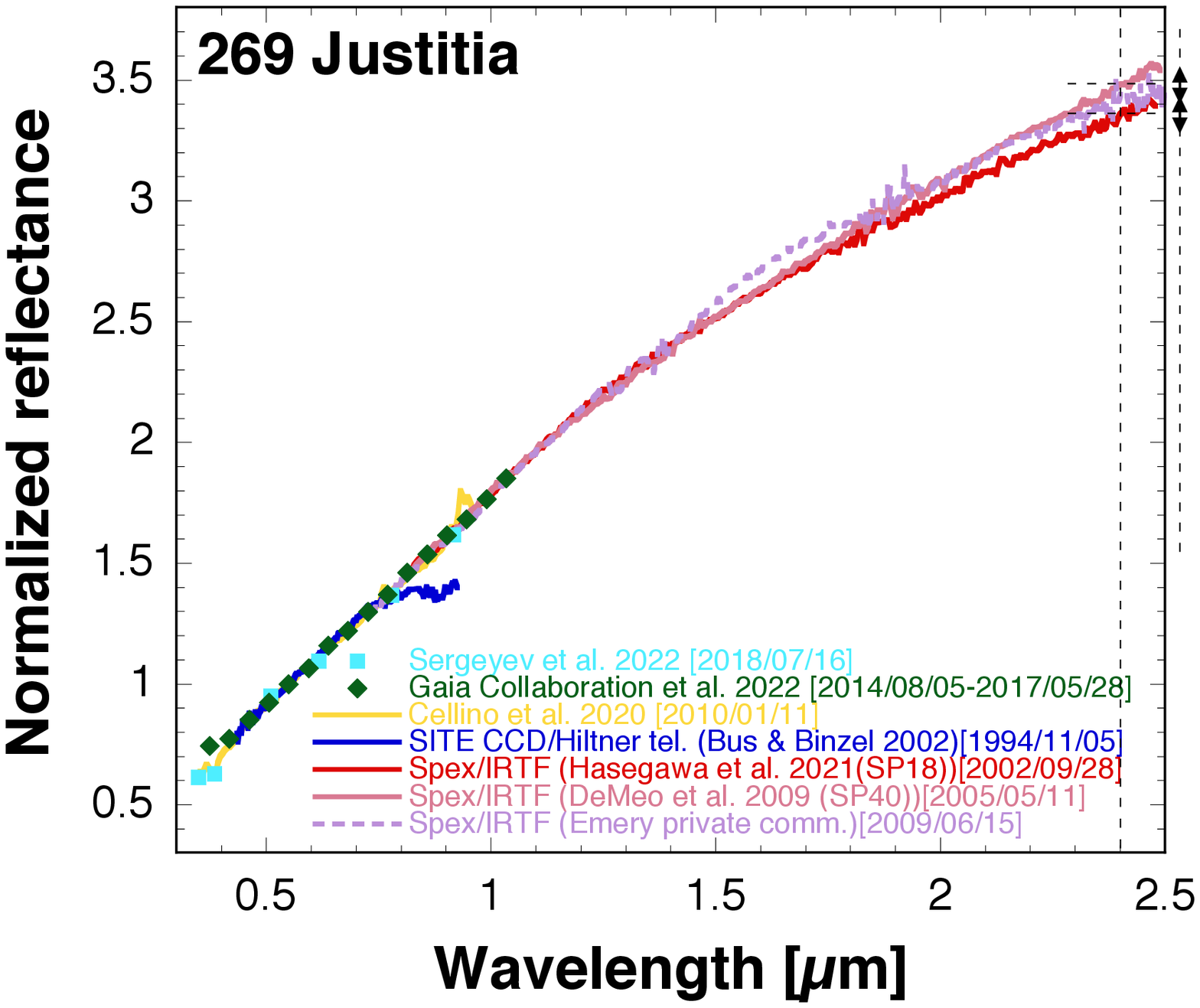}{0.5\textwidth}{}
             \fig{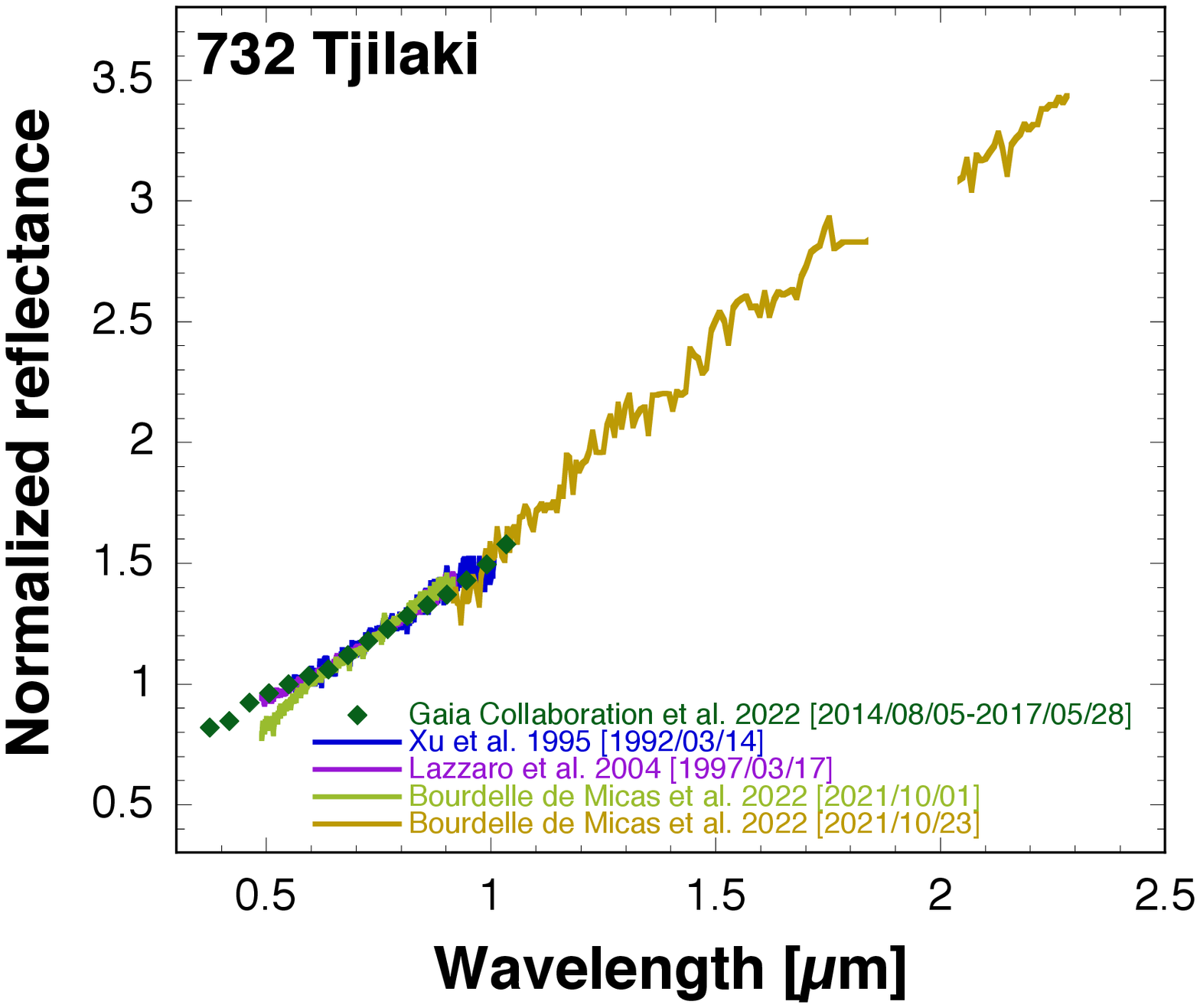}{0.5\textwidth}{}
}
\caption{
Spectra of 203 Pompeja (top panels), 269 Justitia (bottom left panel), and 732 Tjilaki (bottom right panel).
Square brackets in these panels indicate the observation dates.
The black arrows indicate the slope uncertainty from \citet{Marsset2020} centered at the reflectance value of the spectra at 2.40 $\mu$m, as shown by the black dashed lines.
}
\label{fig:203269}
\end{figure*}

The visible spectra of 269 Justitia measured on 1994 November 05 and 2010 January 11 were retrieved from \citet{Bus2002} and \citet{Cellino2020}, respectively.
Most data collected on 2018 July 16 were retrieved from \citet{Sergeyev2022}.
The mean visible spectra was also obtained somewhere between 2014 August 05 and 2017 May 28 \citep{GaiaCollaborationInPress}\textsuperscript{\ref{JPL}}.
These visible spectra of 269 at 0.45--0.85 $\mu$m are consistent with one another (Figure \ref{fig:203269}).
This indicates that the visible spectrum of 269 at 0.45--0.85 $\mu$m did not change over 22 years.
Two NIR spectra of 269 were acquired on 2002 September 28 \citep{Hasegawa2021b} and 2005 May 11 \citep{DeMeo2009}.
Moreover, the NIR spectrum of 269 was obtained on 2009 June 15 (Emery, private comm.).
The slopes of the three NIR data between 0.8--0.95 $\mu$m are consistent with the slope of the visible data from \citet{Sergeyev2022}  (Figure \ref{fig:203269}). 

The slope of the visible spectrum of 269 between 0.45--0.95 $\mu$m is found to have remained unchanged for at least about 16 years.
Although the observed wavelengths are different, the results of unchanged spectra are the same for 269 and 596.

\section{732 Tjilaki} \label{sec:732}
Visible spectra of 732 Tjilaki were obtained on 1992 March 04 \citep{Xu1995}, 1997 March 07 \citep{Lazzaro2004}, and 2021 October 01 \citep{BourdelledeMicasinPress}.
The mean visible spectra was also acquire between 2014 August 05 and 2017 May 28 \citep{GaiaCollaborationInPress}\textsuperscript{\ref{JPL}}.
NIR spectrum was taken on 2021 October 23 were retrieved from \citet{BourdelledeMicasinPress}.
These visible spectra of 732 at 0.6--0.9 $\mu$m are consistent with one another (Figure \ref{fig:203269}).

The slope of the visible spectrum between 0.6--0.9 $\mu$m of 732 has not changed for at least about 29 years.
We conclude that the spectra of 732 have not changed within the observed time interval, which is consistent with our results for 269 or 596.

\section{203 Pompeja} \label{sec:203}
For 203 Pompeja, a visible spectrum, visible spectrophotometric color, and NIR spectrum were taken on 2021 January 29, Febrary 4, and January 14, respectively \citep{Hasegawa2021b}.
Meanwhile, most of the data from \citet{Sergeyev2022} for 203 were acquired from 2018 April 16 to May 15.
The visible spectra of 203, which were acquired by \citet{Sawyer1991,GaiaCollaborationInPress}\textsuperscript{\ref{JPL}}, match the one of \citet{Sergeyev2022}.
The slope of the spectrum of 203 obtained by \citet{Hasegawa2021b} is redder than that of a typical D type, but the slope of the spectrum obtained by \citet{Sergeyev2022} is comparable to that of a T type (Figure \ref{fig:203269}).
In addition, new visible and NIR spectra of 203 were obtained on 2022 June 18, 19, 29, and 30, and 2020 December 1 and 2022 June 10 and 27, respectively and classified as X types.
The above indicate that there is a large slope variation of the visible and NIR spectra of 203 acquired in \citet{Hasegawa2021b} with respect to the other years.
Moreover, the latitude and longitude of the Sub-Earth point at the time of both observations by IRTF coincide (see Table \ref{tab:B1}).
Table \ref{tab:B1} was based on the shape model of 203 from Appendix \ref{sec:203shape}.

\begin{deluxetable*}{llccccccccl}
\label{tab:B1}
\tablecaption{Sub-solar and observer coordinates of 203 Pompeja during photometric and spectral observations}
\tablewidth{0pt}
\tablehead{
\colhead{Observation time}&\colhead{Telescope}&\multicolumn4c{Sub-solar point}&\multicolumn4c{Sub-Earth point}&\colhead{Reference}\\
\colhead{}&\colhead{}&\colhead{${\lambda_{1}}^{a}$ [\degr]}&\colhead{${\phi_{1}}^{a}$ [\degr]}&\colhead{${\lambda_{2}}^{b}$ [\degr]}&\colhead{${\phi_{2}}^{b}$ [\degr]}&\colhead{${\lambda_{1}}^{a}$ [\degr]}&\colhead{${\phi_{1}}^{a}$ [\degr]}&\colhead{${\lambda_{2}}^{b}$ [\degr]}&\colhead{${\phi_{2}}^{b}$ [\degr]}&\colhead{}
}
\startdata
1988/10/06 08:15&McDonald&60&$-$1&234&2&51&11&225&$-$10&\cite{Sawyer1991}\\
2018/04/16 14:37--54&SkyMapper&21--16&$-$4&202--197&4&15--10&$-$13&194--190&11&\cite{Sergeyev2022}\\
2018/04/17 13:55&SkyMapper&32&$-$4&212&4&26&$-$13&205&11&\cite{Sergeyev2022}\\
2018/04/21 14:14--30&SkyMapper&30--26&$-$5&211--207&4&25-21&$-$13&204-200&311&\cite{Sergeyev2022}\\
2018/04/22 13:42&SkyMapper&39&$-$5&219&5&34&$-$13&213&11&\cite{Sergeyev2022}\\
2018/05/15 14:11--12&SkyMapper&47&$-$9&228&8&49&$-$9&228-227&7&\cite{Sergeyev2022}\\
2019/09/01 13:36&SkyMapper&334&$-$48&156&49&340&$-$50&162&50&\cite{Sergeyev2022}\\
2019/09/07 09:48&SkyMapper&34&$-$48&216&48&44&$-$50&226&50&\cite{Sergeyev2022}\\
2020/12/01 08:21&IRTF&332&29&153&$-$28&337&35&156&$-$32&This study\\
2021/01/14 09:08&IRTF&357&36&179&$-$36&9&29&189&$-$26&\cite{Hasegawa2021b}\\
2021/01/29 12:55&SAO(Spec)&310&39&132&$-$38&326&28&146&$-$26&\cite{Hasegawa2021b}\\
2021/02/04 10:36&SAO(Phot)&348&39&170&$-$39&5&27&189&$-$26&\cite{Hasegawa2021b}\\
2022/06/10 05:41&IRTF&341&17&163&$-$17&357&31&175&$-$33&This study\\
2022/06/18 23:27&TRAPPIST&80&16&262&$-$16&97&30&277&$-$32&This study($BVR_{\rm C}I_{\rm C}$)\\
2022/06/19 01:02&TRAPPIST&57&16&239&$-$16&73&30&253&$-$32&This study($BVR_{\rm C}I_{\rm C}$)\\
2022/06/19 02:43&TRAPPIST&32&16&213&$-$16&48&30&228&$-$32&This study($BVR_{\rm C}I_{\rm C}$)\\
2022/06/27 08:08&IRTF&316&14&138&$-$15&332&29&152&$-$31&This study\\
2022/06/29 00:18&TRAPPIST&75&14&257&$-$15&91&29&271&$-$31&This study($BVR_{\rm C}I_{\rm C}z'$)\\
2022/06/29 23:15&TRAPPIST&91&14&273&$-$14&107&28&287&$-$31&This study($BVR_{\rm C}I_{\rm C}z'$)\\
2022/06/30 00:57&TRAPPIST&66&14&248&$-$14&82&28&262&$-$30&This study($BVR_{\rm C}I_{\rm C}z'$)\\
2022/06/30 02:30&TRAPPIST&43&14&225&$-$14&58&28&239&$-$30&This study($BVR_{\rm C}I_{\rm C}$)\\
\enddata
\tablecomments{
\\ 
$^{a}$Pole solution 1: ${\lambda}_1$ = 315 [\degr], ${\beta}_1$ = $-$40 [\degr], Period = 24.0550 [hr]\\ 
$^{b}$Pole solution 2: ${\lambda}_2$ = 132  [\degr], ${\beta}_2$ = $-$36  [\degr], Period = 24.0551 [hr]\\
}
\end{deluxetable*}

The fact that the NIR spectra of other asteroids obtained using the same solar analogue stars on the same date as \citet{Hasegawa2021b} are consistent with previous spectra gives credit to this NIR spectrum of 203 (see Appendix \ref{sec:203-old}).
Additional visible data for three other asteroids obtained by the same surveys as 203 are highly consistent with the data from \citet{Sergeyev2022} and \citet{GaiaCollaborationInPress} (see Appendix \ref{sec:SNU-obs}). 
Therefore, it appears that the visible spectrum and spectrophotometric color reported in \citet{Hasegawa2021b} are valid.
Furthermore, the slope of the measured visible spectrum (0.7--0.8 $\mu$m) is consistent with the slope in the NIR spectrum (0.8--0.9 $\mu$m), which further validates the observational result of \citet{Hasegawa2021b}.

The first possible cause of the spectral differences is cometary activity, as reported on 596.
Therefore, the radial profiles of 203 around the observation period of \citet{Hasegawa2021b} were investigated by comparing the radial profiles with field stars to detect a faint comet-like coma.
However, no coma is detected within the background detection limit of 25.1 mag arcsec$^{-2}$ in the $r$ band (see Appendix \ref{sec:203coma}).
This evidence suggests that the spectral differences may not be triggered by an unforeseen cometary activity.

The only difference between the IRTF observations is the latitude of the sub-solar point.
Sub-solar points were at relatively high latitudes\footnote{This is applied to the first shape/pole model, whereas the second shape/pole solution has the northern and southern hemispheres reversed.\label{203shape}} in 2021 observations (Table \ref{tab:B1}).
The existence of regions with a VRO-like spectrum in high latitudes is necessary to explain the spectral differences.
There may be a giant crater, that is shaded under low sub-solar point and whose areas are not visible, such as the giant crater on asteroid  253 Mathilde \citep{Veverka1997}, for example.
Since the VRO-like spectra are much brighter in the NIR than the X-type spectra, even if the area of the crater illuminated by sunlight is small, the feature may appear in the spectrum.
An asteroid with surface heterogeneity was also reported in \citet{Avdellidou2021}.
These authors suggested that the very red spectrum may correspond to a fresh crater formed by an impact as their studied asteroid 223 Rosa, also showed a very red spectrum when observed at different epochs.
Finding craters with a VROs-like color on asteroids with an X-type surface implies that asteroids with a VROs surface layer evolved to an X-type-like surface layer by space weathering.
If solar wind is the main cause of space weathering, the time of sunlight exposure at the bottom of craters such as this one, where sunlight is less directly illuminated, is much shorter than on other surfaces, so that the time for sunlight penetration is relatively longer and the spectral features may not disappear.
Considering that VROs have the same spectral properties as RR and IR class trans-Neptunian objects \citep{Fulchignoni2008,FraserSubmited}, this supports the proposal of \citep{Vernazza2021,Rivkin2022} that D/T-type and dark X-complex asteroids originated from the same transplanetary and/or transneptunian planetesimal region.

\section{Conclusions} \label{sec:Conclusions}
A comparative study of the spectra of the T/D-type asteroid 596 Scheila before, immediately after, and 10 years after the 2010 impact event reveals the following:

 \begin{itemize}
\item Fresh surfaces on dark asteroids with reddish spectra like 596 do not undergo any noticeable spectral changes induced by space weathering over a 10-year timescale.
\item This is the first time that the absence of spectral changes due to space weathering has been measured over an accurate time interval for an actual asteroid, as opposed to meteorite samples measured in laboratory simulations.
\item During the first decade of space weathering following a resurfacing event, the spectral change progression of dark and reddish asteroids does not appear to progress logarithmically but linearly.
\end{itemize}

Applying the findings of our study to VROs: 203 Pompeja, 269 Justitia, and 732 Tjilaki, we can suggest the following:

 \begin{itemize}
\item 
The spectrum of 203 Pompeja was found to be a VRO in the 2021 observations, but an X type in the other observational epoch.
The existence of a giant crater with a VRO spectrum in one of the polar regions is suggested to explain this spectral difference.
This would imply that the spectra of VROs evolve into dark X types as space weathering progresses over time.
\item Combining the results from the study of \citet{Fatka2022} on asteroid pairs and our own results, it is possible that the spectral slope of VROs remains mostly unchanged over a timescale of 300 years.
\end{itemize}

We would like to thank the referee for the careful review and constructive suggestions, which helped us to improve the manuscript significantly.  
We are grateful to Dr. Joshua P. Emery for sharing valuable 269 data and supporting 203 observation.
We greatly appreciate Dr. Petr Fatka and Dr. Nicholas A. Moskovitz for useful information about their observed NEAs. 
This work is based on observations collected at the Infrared Telescope Facility, which is operated by the University of Hawaii under contract 80HQTR19D0030 with the National Aeronautics and Space Administration, Ishigakijima Astronomical Observatory, National Astronomical Observatory of Japan, and at Seoul National University Astronomical Observatory. 
The authors acknowledge the sacred nature of Mauna kea and appreciate the opportunity to observe from the mountain. 
The instrumentation at IAO was supported by a Grant-in-Aid for Scientific Research on Priority Areas (19047003).
TRAPPIST is a project funded by the Belgian Fonds (National) de la Recherche Scientique (F.R.S.-FNRS) under grant PDR T.0120.21.
We thank the Las Cumbres Observatory and their staff for its continuing support of the ASAS-SN project. 
ASAS-SN is supported by the Gordon and Betty Moore Foundation through grant GBMF5490 to the Ohio State University, and funded in part by the Alfred P. Sloan Foundation grant G-2021-14192 and NSF grant AST-1908570. 
Development of ASAS-SN has been supported by NSF grant AST-0908816, the Mt. Cuba Astronomical Foundation, the Center for Cosmology and AstroParticle Physics at the Ohio State University, the Chinese Academy of Sciences South America Center for Astronomy (CAS-SACA), the Villum Foundation, and George Skestos.
This work has made use of data from the Asteroid Terrestrial-impact Last Alert System (ATLAS) project. 
ATLAS is primarily funded to search for near earth asteroids through NASA grants NN12AR55G, 80NSSC18K0284, and 80NSSC18K1575; byproducts of the NEO search include images and catalogs from the survey area.  
The ATLAS science products have been made possible through the contributions of the University of Hawaii Institute for Astronomy, the Queen's University Belfast, the Space Telescope Science Institute, the South African Astronomical Observatory (SAAO), and the Millennium Institute of Astrophysics (MAS), Chile.
ZTF data were based on observations obtained with the Samuel Oschin Telescope 48-inch and the 60-inch Telescope at the Palomar Observatory as part of the Zwicky Transient Facility project. 
ZTF is supported by the National Science Foundation under GrantNo. AST-2034437 and a collaboration including Caltech, IPAC, the Weizmann Institute for Science, the Oskar Klein Center atStockholm University, the University of Maryland, Deutsches Elektronen-Synchrotron and Humboldt University, the TANGOConsortium of Taiwan, the University of Wisconsin at Milwaukee, Trinity College Dublin, Lawrence Livermore NationalLaboratories, and IN2P3, France. Operations are conducted by COO, IPAC, and UW.
This study has utilized the SIMBAD database, operated at CDS, Strasbourg, France, and the JPL HORIZONS ephemeris generator system, operated at JPL, Pasadena, USA. 
This work is based on data provided by the Minor Planet Physical Properties Catalogue (MP3C) of the Observatoire de la C\^{o}te d'Azur.
F.D. and M.M. were supported by the National Aeronautics and Space Administration under grant No. 80NSSC18K0849 and 80NSSC18K1004 issued through the Planetary Astronomy Program.
Any opinions, findings, and conclusions or recommendations expressed in this letter are those of the authors and do not necessarily reflect the views of the National Aeronautics and Space Administration.
The work of J.H. has been supported by the Czech Science Foundation through grant 20-08218S.
C.A. and M.D. acknowledge support from ANR "ORIGINS" (ANR-18-CE31-0014).  
C.A. and M.D. were Visiting Astronomers at the Infrared Telescope Facility, which is operated by the University of Hawaii under contract 80HQTR19D0030 with the National Aeronautics and Space Administration.
E.J. is FNRS Senior Research Associate.
M.Im acknowledges the support from the Korea Astronomy and Space Science Institute grant under the R\&D program (Project No.2020-1-600-05) supervised by the Ministry of Science and Technology and ICT (MSIT), and the National Research Foundation of Korea (NRF) grant, No. 2020R1A2C3011091, funded by MSIT. 
M.I. was supported by the NRF grant No. 2018R1D1A1A09084105.
This study was supported by JSPS KAKENHI (grant nos. JP20K04055, JP21H01140, JP21H01148, and JP22H00179) and by the Hypervelocity Impact Facility (former facility name: the Space Plasma Laboratory), ISAS, JAXA.

\appendix
\section{Shape model of 203 Pompeja} \label{sec:203shape}
In order to link the measured spectra of 203 Pompeja to their corresponding regions on its surface (i.e., to derive the sub-observer points), the knowledge of the rotation state (i.e., sidereal rotation period and orientation of the spin axis) is necessary. 
However, 203's shape model and rotation state had not been derived so far. 
To fix this, we compiled available optical data of 203 and applied the convex inversion. 
This light curve inversion method developed by \citet{Kaasalainen2001a, Kaasalainen2001b} was already successfully used for the 3D shape and rotation state determination of more than 3,400 asteroids (see the Database of Asteroid Models from Inversion Techniques, DAMIT\footnote{\url{https://astro.troja.mff.cuni.cz/projects/damit/}}, \citealt{Durech2010}). 
Therefore, convex inversion is a well tested state-of-the art method for the shape modeling of asteroids. 
Here we follow the standard modeling approach described in details, for instance, in \citet{Hanus2021, AthanasopoulosInPress}.

Our optical dataset include 18 dense light curves from two apparitions: in 1983 \citep{DiMartino1984b} and 2011/2012 \citep{Pilcher2012h}. 
Especially the latter dataset is highly valuable as the authors dedicated a large effort in reliably deriving the rotation period of 203. 
The main difficulty stems from the rotation period being very close to 24 hr, the rotation phase then cannot be fully sampled from one location on the Earth, but rather multiple observers at different longitudes have to participate in the campaign. 

Furthermore, we supplemented the dense light curves with the sparse-in-time photometric measurements from five different sky-surveys. 
In particular, we used data from the Catalina Sky Survey \citep[CSS;][]{Larson2003}, the US Naval Observatory in Flagstaff (USNO-Flagstaff), the Asteroid Terrestrial-impact Last Alert System \citep[ATLAS;][]{Tonry2018, Tonry2018b, Heinze2018}, the All-Sky Automated Survey for Supernovae \citep[ASAS-SN;][]{Shappee2014,Kochanek2017,Hanus2021}, and the Gaia Data Release 2 \citep[Gaia DR2;][]{Spoto2018}.

\begin{figure*}
\gridline{\fig{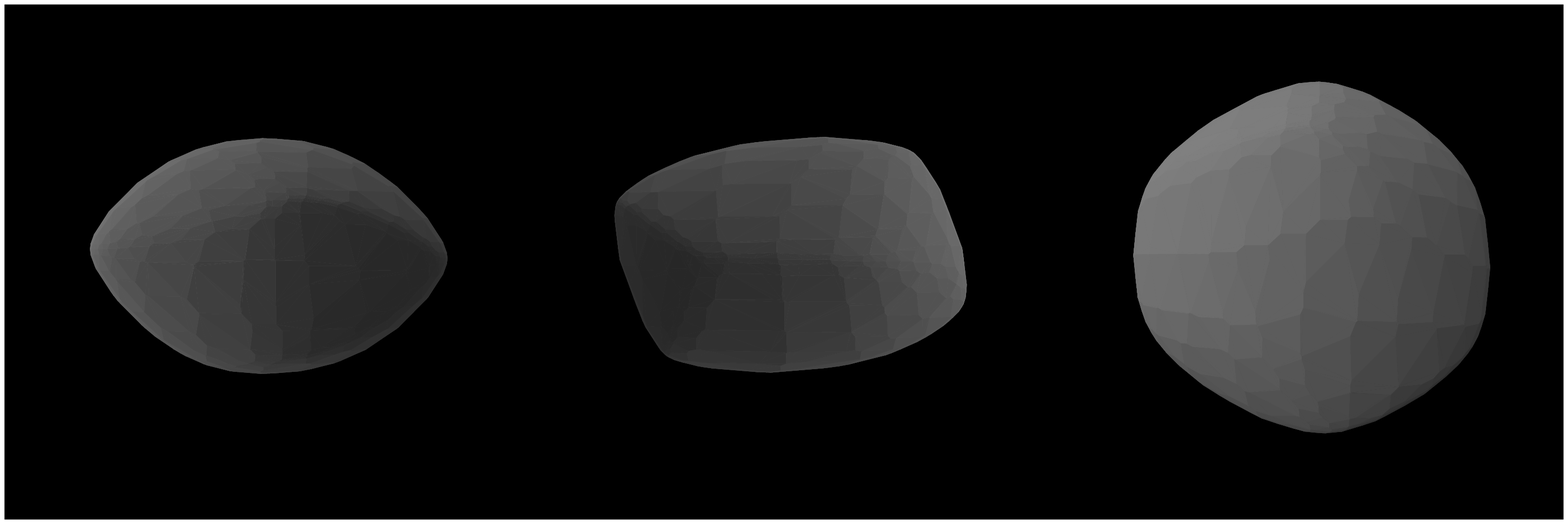}{0.5\textwidth}{}
          \fig{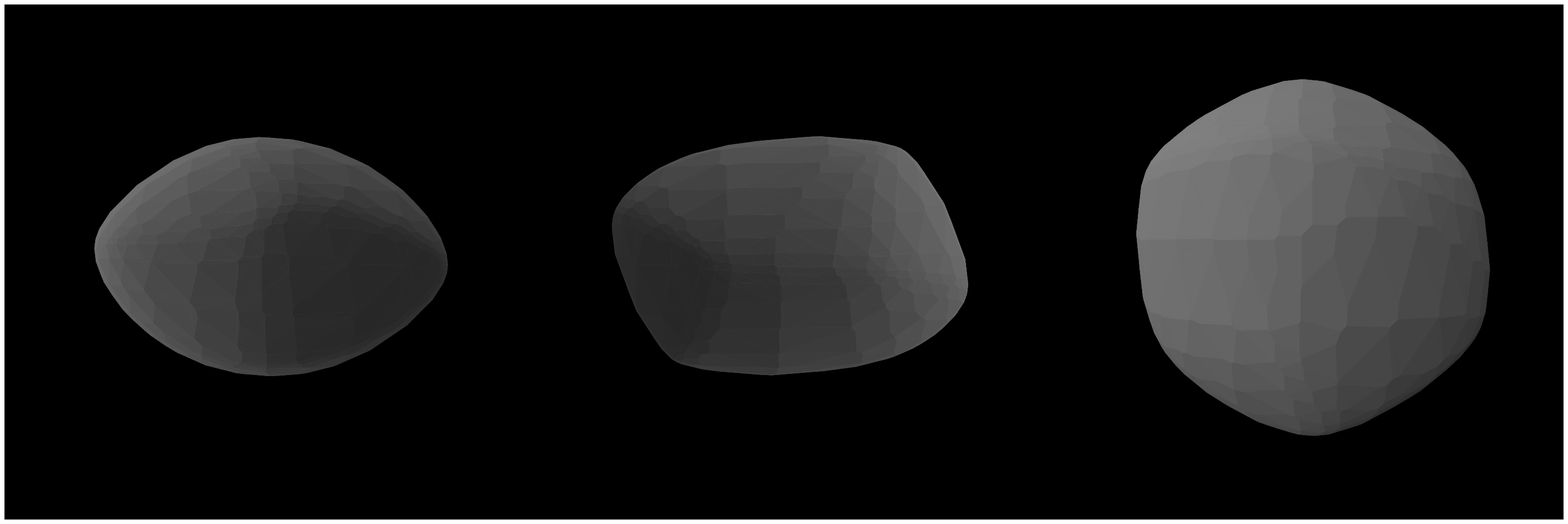}{0.5\textwidth}{}
          }
\caption{Convex shape models of 203 Pompeja. 
While the left panel represents the pole solution ${\lambda}_1$ = 315 [\degr], ${\beta}_1$ = $-$40 [\degr], the right panel shows the pole solution ${\lambda}_2$ = 132  [\degr], ${\beta}_2$ = $-$36  [\degr]. 
The left and center views are equatorial with a 90$^\circ$ rotation, while the right view is pole-on. Both shape models appear very similar.
}
\label{fig:shape}
\end{figure*}

The optical dataset for 203 allowed us to derive its unique spin state and 3D shape model solution. 
203 rotates with a sidereal rotation period of (24.05502 $\pm$ 0.00005)~hr and we provide two symmetric pole solutions following the standard ambiguity of the inversion method \citep{Kaasalainen2004}.  
The appearance of the two shape models in shown in Fig.~\ref{fig:shape}.

\restartappendixnumbering

\section{A NIR observation of 203 Pompeja in 2021 January} \label{sec:203-old}
\begin{figure*}
\gridline{\fig{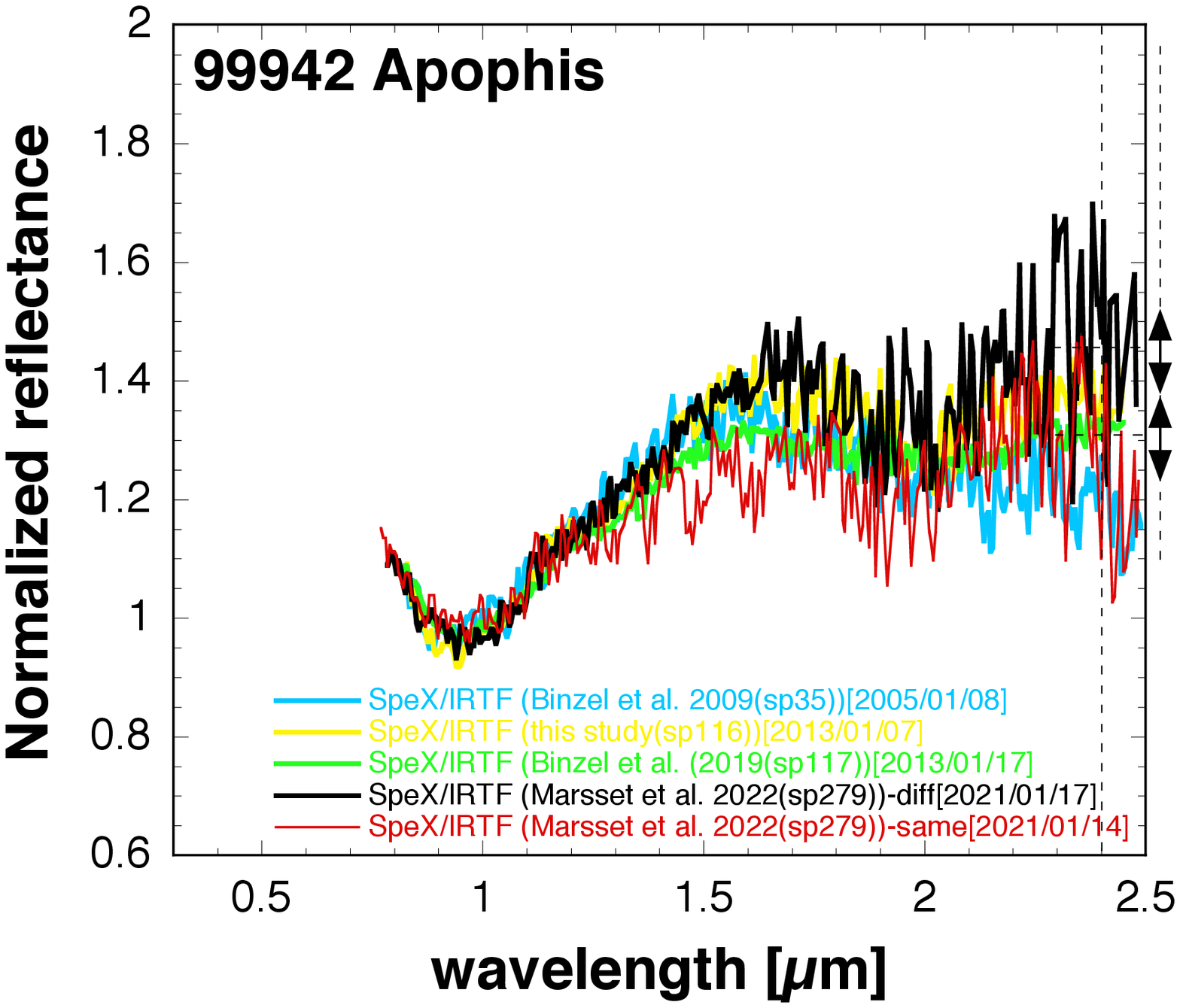}{0.5\textwidth}{}
          \fig{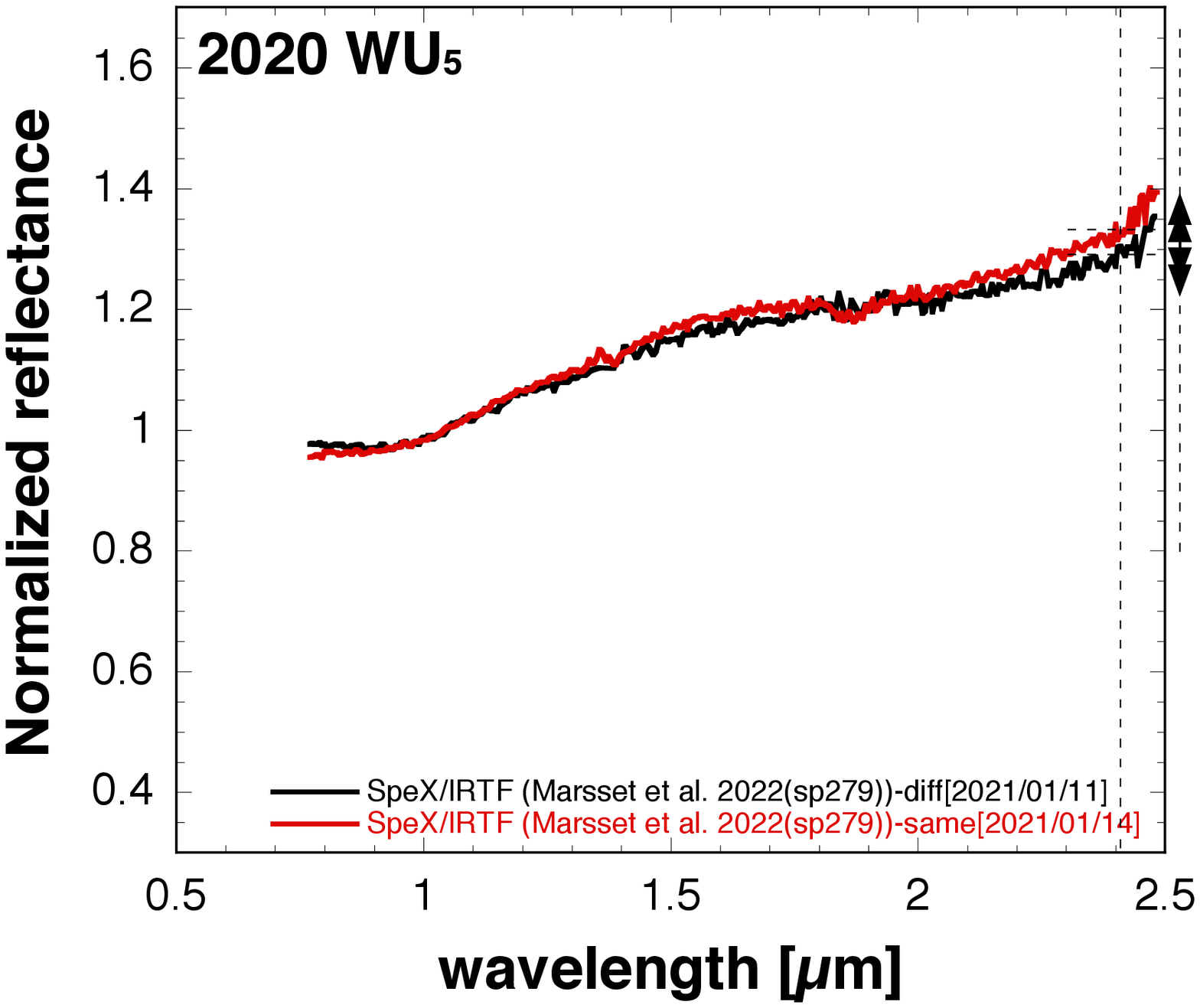}{0.5\textwidth}{}
          }
\caption{
Spectra of 99942 Apophis  (left panel) and 2020 $\mathrm{WU_{5}}$ (right panel).
Data of 203 Pompeja, 99942 Apophis, and 2020 $\mathrm{WU_{5}}$ taken on 2021 January 14. 
The spectra of 99942 and 2020 $\mathrm{WU_{5}}$ were measured four and one times, respectively \citep{Binzel2009,Binzel2019,Marsset2022}, in addition to their measurement on the same day as 203 \citep{Hasegawa2021b}.
The five spectra of 99942 and the two spectra of 2020 $\mathrm{WU_{5}}$ are found to be in agreement.
Square brackets in these panels indicate the observation dates.
The black arrows indicate the slope uncertainty from \citet{Marsset2020} centered at the reflectance value of the spectra at 2.40 $\mu$m, as shown by the black dashed lines.
}
\label{fig:99942}
\end{figure*}

The spectrum of 203 Pompeja measured on 2021 January 14 \citep{Hasegawa2021b} is much redder than usual asteroids, which may cast doubt on the validity of the observations.
One possibility could be that the spectrum was red due to slit misalignment.
However, if the asteroid is out of alignment, the spectrum is  bluing instead of reddening \citep{Marsset2020}.
Therefore, in order for the asteroid to become red, it is the solar analogue star observations that needed to be misaligned.
On the night we observed 203 (2021 January 14), we also observed five solar analogue stars for the spectral division, four of which were are spectrally very consistent (the fifth star was discarded).
In addition, we observed NEAs 99942 Apophis and 2020 $\mathrm{WU_{5}}$ on the same night as 203 and their respective spectra were consistent with previous and more recent observations (Figure \ref{fig:99942}).
This provide strong credit to our 2021 observations of 203.
Future spectroscopic observations  and high-angular resolution imaging of 203 acquired under high sub-observer latitudes will help confirming or dismissing the existence of a fresh polar crater with a VRO spectrum on 203.

\section{Visible observations of three asteroids after 2020} \label{sec:SNU-obs}
Spectroscopic and spectrophotometric observations of three MBAs: 212 Medea, 328 Gudrun, and 769 Tatjana were performed with the Shelyak PF0021VIS--LISA spectrometer and the SBIG STX-16803 camera equipped with $V$-, $R_{\rm C}$-, and $I_{\rm C}$-band filters on the 1.0 m telescope at Seoul National University Astronomical Observatory (SAO), Seoul, Republic of Korea \citep{Im2021} in 2020 or later.
The procedures for the visible observations and reduction are as described in \citet{Hasegawa2021b}.

\begin{figure*}
\gridline{\fig{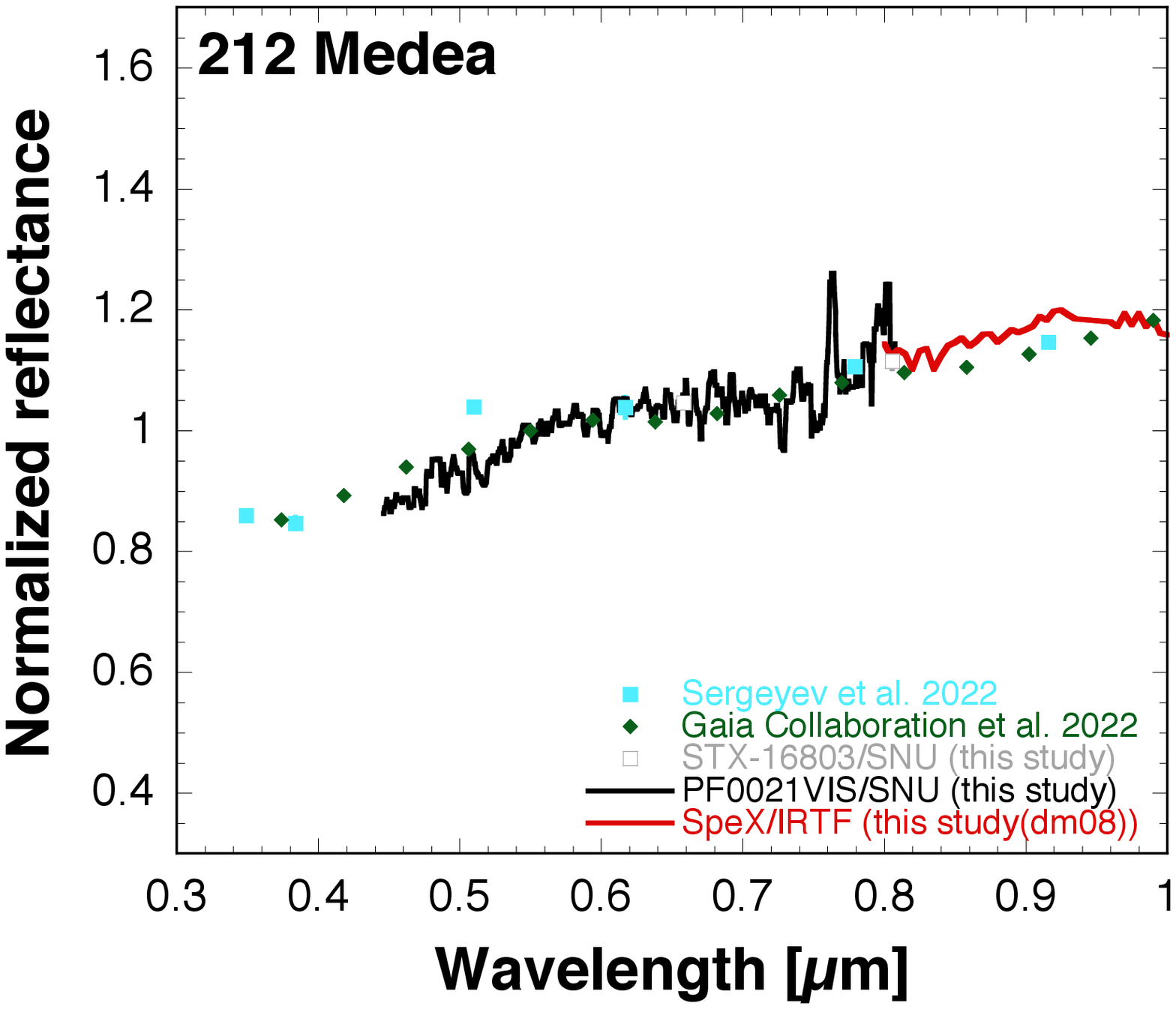}{0.33\textwidth}{}
          \fig{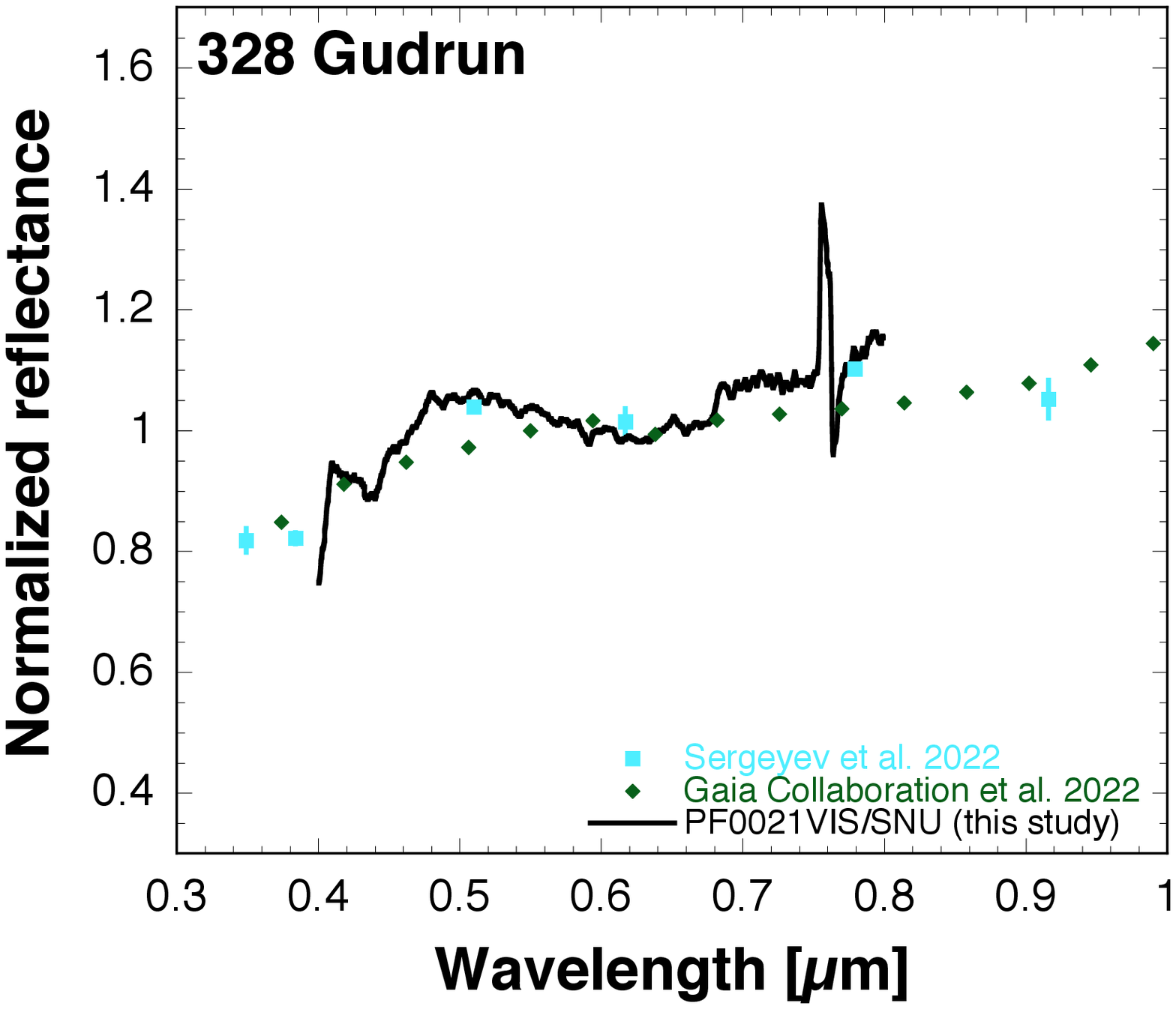}{0.33\textwidth}{}
          \fig{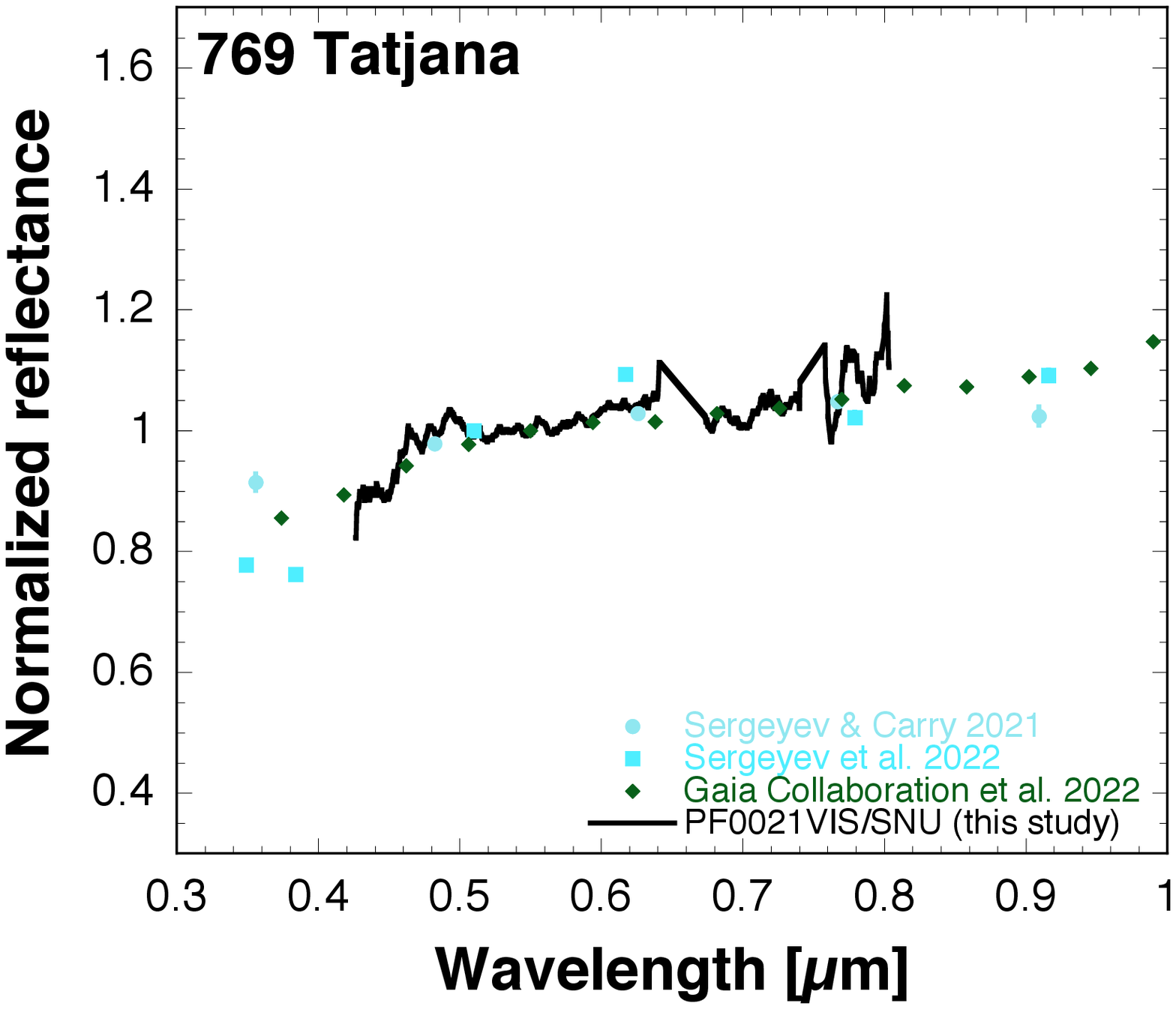}{0.33\textwidth}{}
          }
\caption{Spectra of 212 Medea, 328 Gudrun, and 769 Tatjana. 
Spectra of these asteroids were obtained at SAO in 2020--2021.
Previous spectra were derived from \citet{Sergeyev2021, Sergeyev2022, GaiaCollaborationInPress}.
}
\label{fig:SNU}
\end{figure*}

The spectroscopic data of this study yield classification of 212 Medea, 328 Gudrun, and 769 Tatjana as X, Cgh, and Xc types, respectively. 
The acquired spectra match previously obtained spectra (Figure \ref{fig:SNU}).
Therefore, it appears that the visible spectrum and spectrophotometric color reported in \citet{Hasegawa2021b} are valid.

\restartappendixnumbering

\section{Search for a Coma of 203 Pompeja} \label{sec:203coma}
\begin{figure*}
\gridline{\fig{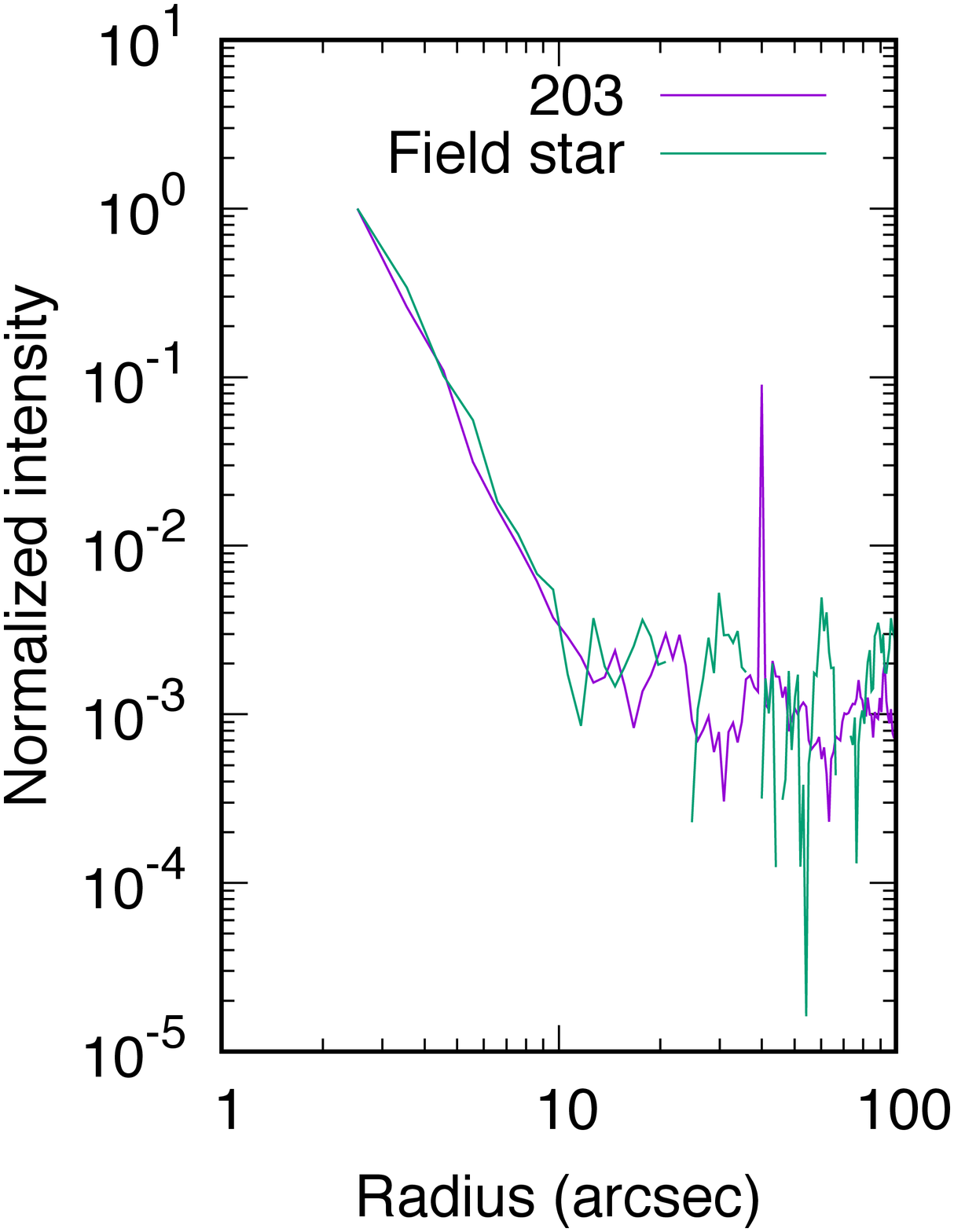}{0.33\textwidth}{}
          \fig{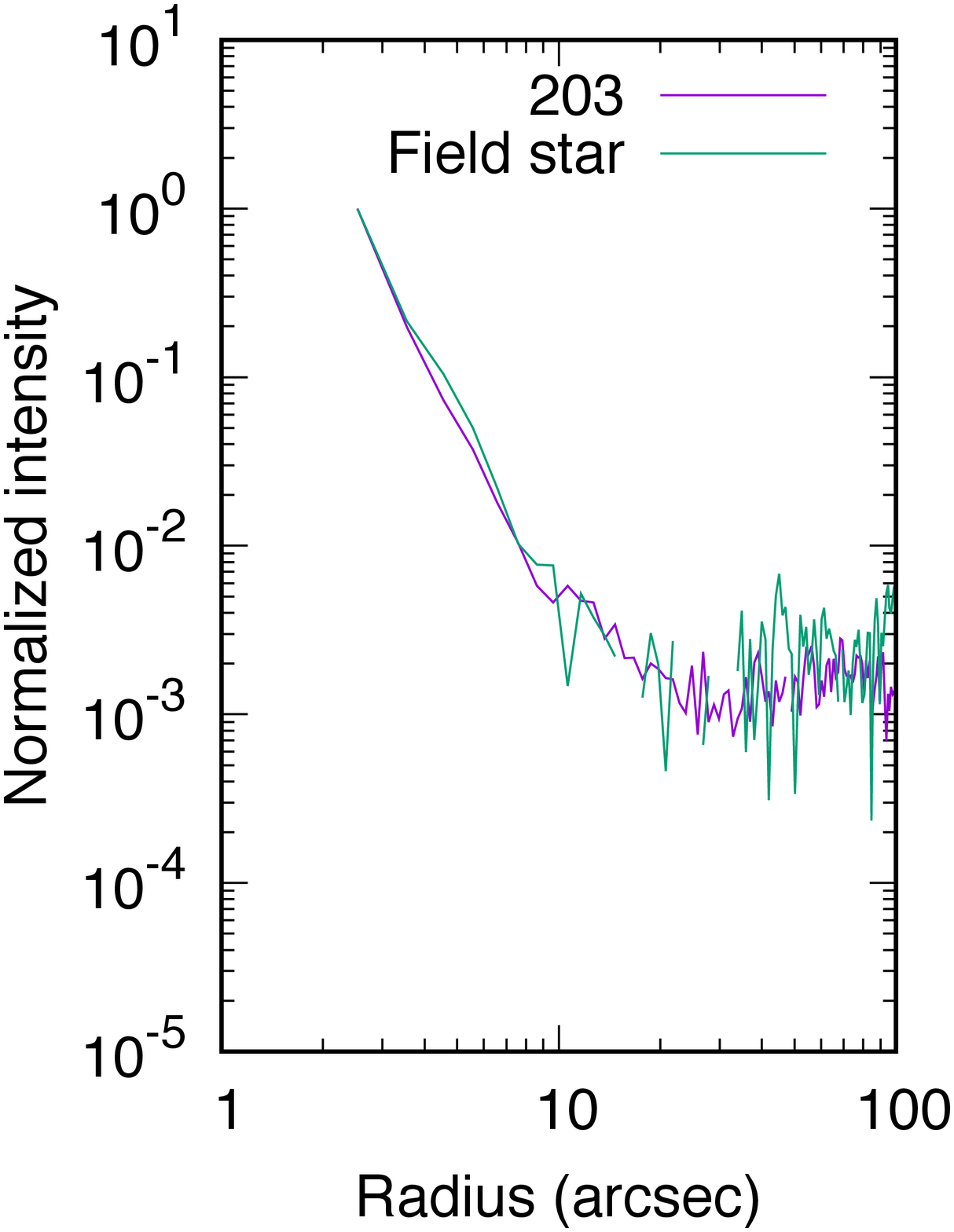}{0.33\textwidth}{}
          \fig{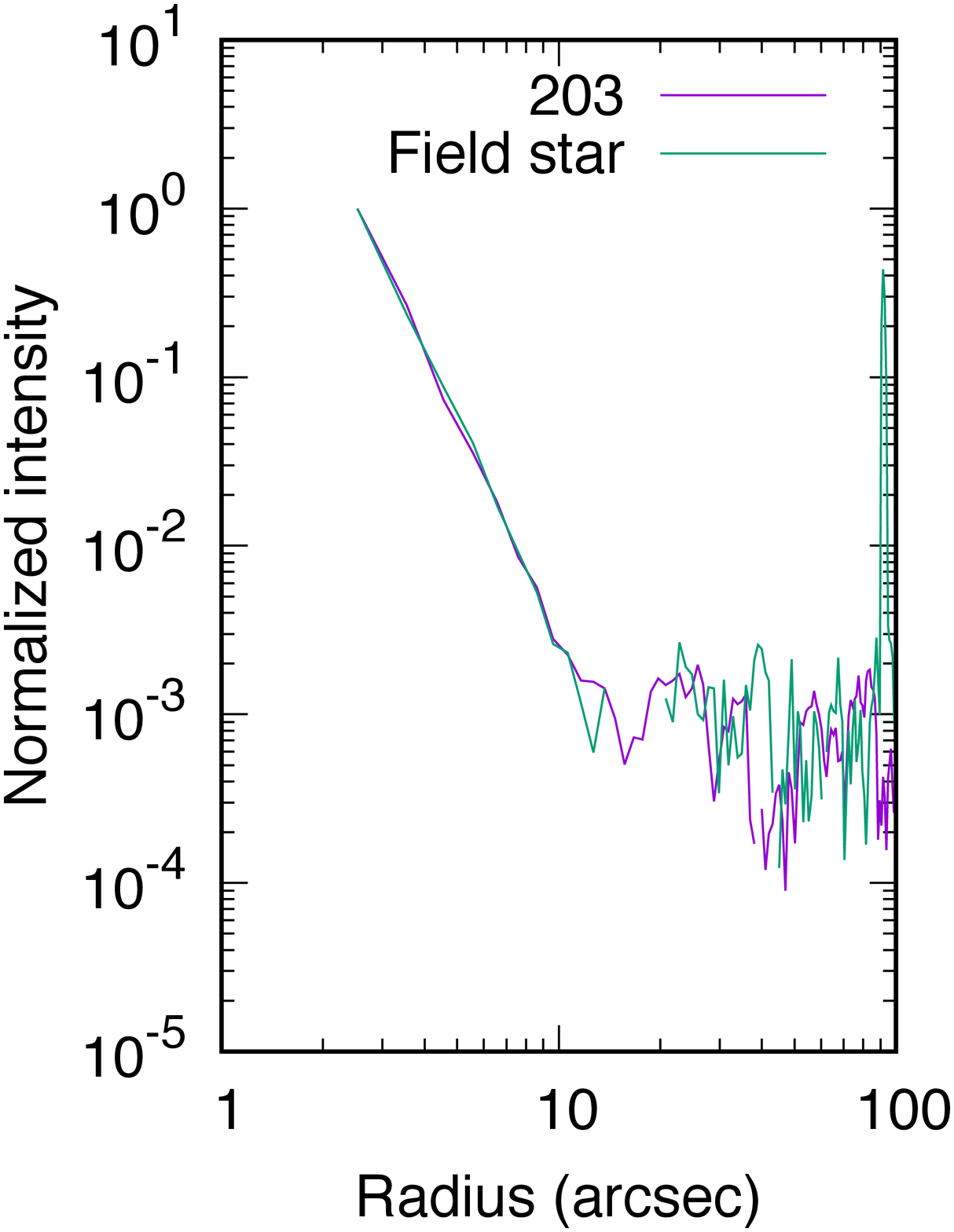}{0.33\textwidth}{}
          }
\gridline{\fig{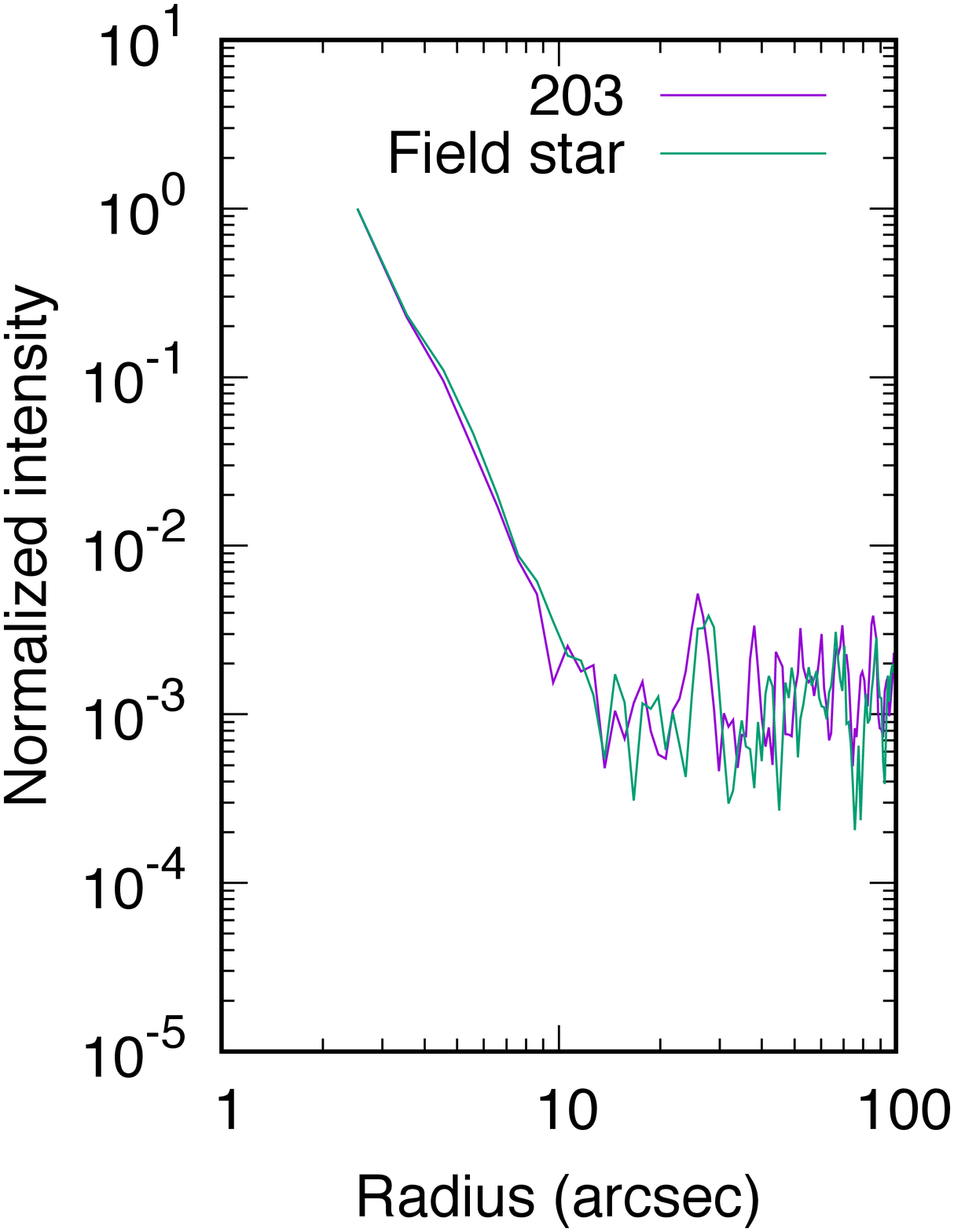}{0.33\textwidth}{}
          \fig{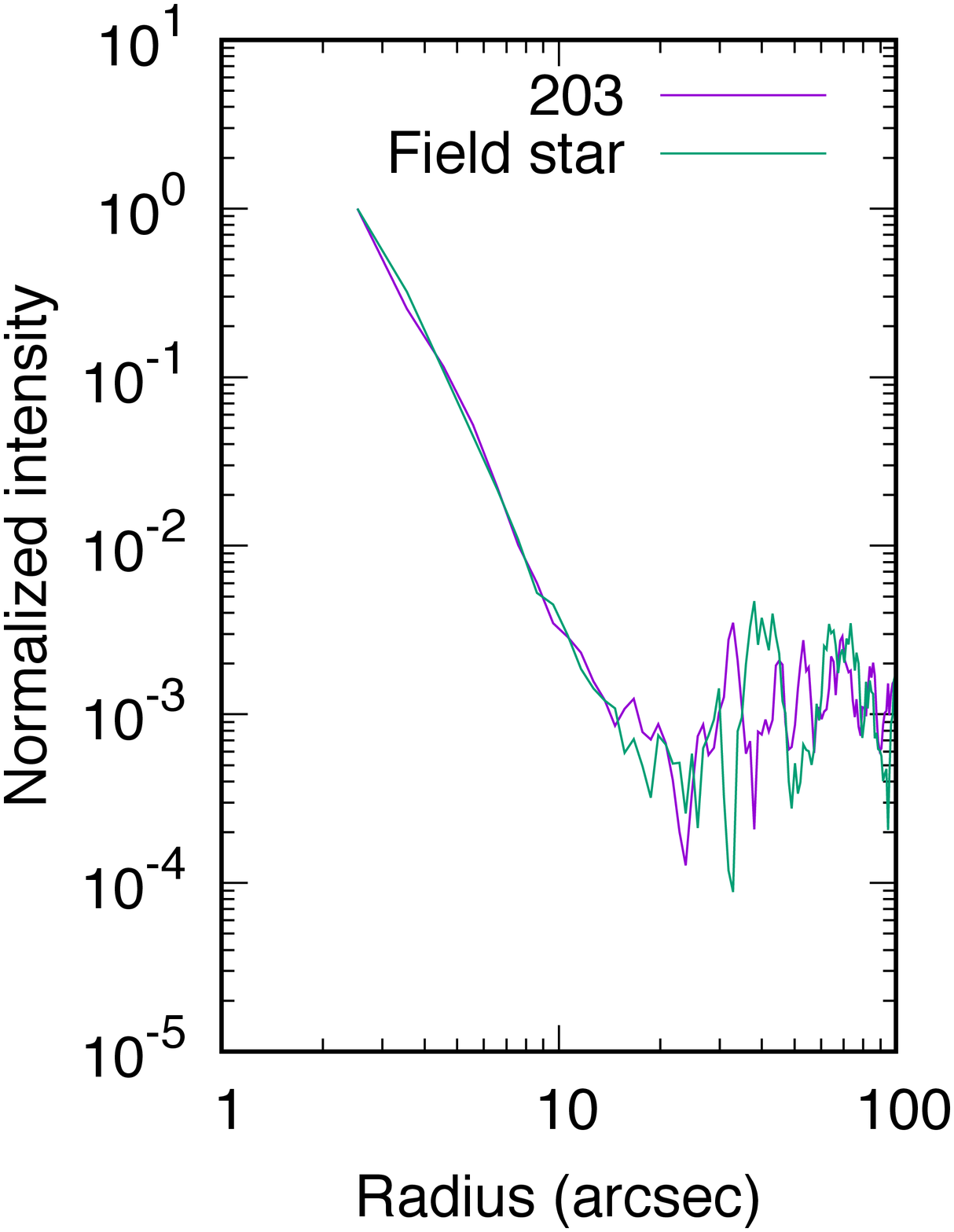}{0.33\textwidth}{}
          \fig{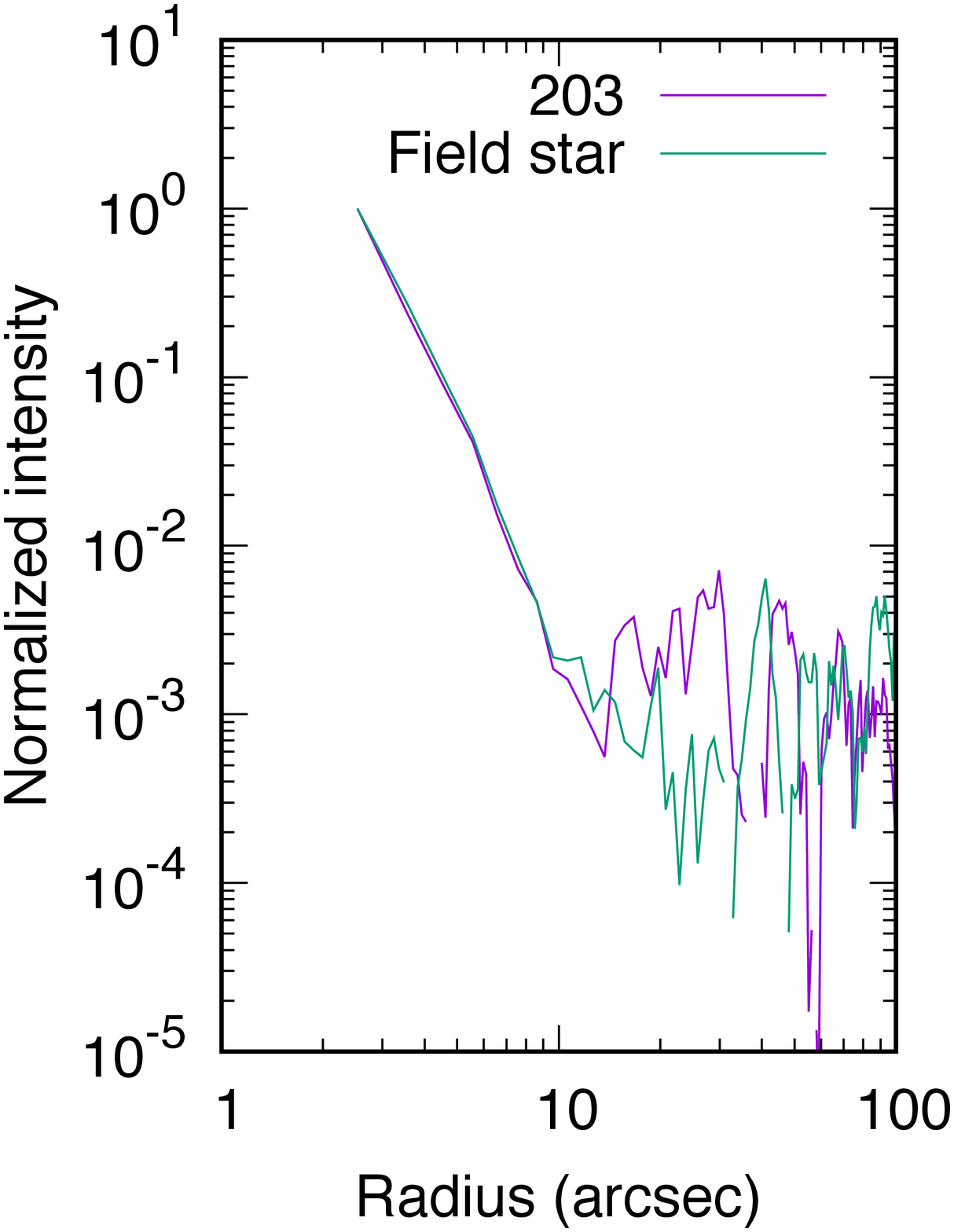}{0.33\textwidth}{}
          }
\caption{
Radial brightness profiles of 203 Pompeja obtained by the Zwicky Transient Facility on 2020 December 10 (top left), 17 (top middle), 22 (top right) and 2021 January 5 (bottom left), 13 (bottom middle), and 19 (bottom right).
These profiles are normalized at 2.5" from the photo center and compared with field stars. 
The FWHM of the images used was roughly 2"--3".
Since 203 was almost saturated, brightness profiles within 2.5" from the photo center are not given. 
}
\label{fig:PSF}
\end{figure*}

To confirm the comet-like activity of 203 Pompeja, radial brightness profiles of 203 and field stars around the observation date of \citet{Hasegawa2021b} were examined. 
For comparison, images in the $r$ band acquired at the 48-inch (1.2 m) Samuel Oschin Schmidt telescope in the Zwicky Transient Facility \citep{Masci2019} were utilized\footnote{\url{https://www.ipac.caltech.edu/doi/irsa/10.26131/IRSA539}\label{IPAC-ZTF}}.
The radial brightness profiles of 203 and field stars are consistent over arange of about three orders of magnitude, and no comet-like coma was detected around the observation date of \citet{Hasegawa2021b} (Figure \ref{fig:PSF}).

\restartappendixnumbering

\facilities{IRTF:3.2m, Murikabushi:1.05m, SAO:1.0m, TRAPPIST-South:0.6m, ZTF:1.2m}


\software{IRAF \citep{Tody1993}, IDL}



\bibliography{hase596R}



\end{document}